\documentclass[sigconf]{acmart} 

\setcopyright{acmlicensed}
\copyrightyear{2024}
\acmYear{2024}

\renewcommand\footnotetextcopyrightpermission[1]{}

\usepackage{float}
\usepackage{array}
\usepackage{rotating}  
\usepackage{pdflscape}
\usepackage{tabularx, multirow, makecell}

\usepackage{adjustbox}
\usepackage{amsmath,amssymb,amsfonts}
\usepackage{algorithmic}
\usepackage[linesnumbered,ruled,vlined]{algorithm2e}
\usepackage{setspace}
\usepackage{graphicx}
\usepackage{textcomp}
\usepackage{xcolor}
\usepackage{multirow}
\usepackage{mdframed}
\usepackage{balance}
\usepackage{booktabs}
\usepackage{tcolorbox}
\usepackage{threeparttable}
\usepackage{colortbl}
\usepackage{subfigure}
\usepackage{hhline}
\usepackage{pifont}
\usepackage{listings}
\usepackage{enumitem}
\usepackage{geometry}
\usepackage{verbatim}
\usepackage{xspace}
\usepackage{listings}
\newcolumntype{C}{>{\centering\arraybackslash}X}
\usepackage{pgfplots}
\pgfplotsset{compat=1.16}
\usetikzlibrary{pgfplots.statistics}

\newcommand{\parabf}[1]{\noindent\textbf{#1}}

\definecolor{ggray}{HTML}{eff0f0}
\definecolor{gggray}{HTML}{E8E8E8}
\definecolor{ggggray}{HTML}{BEBEBE}

\newcommand{\app}{RobGen\xspace}

\newcommand{\qwencoder}{Qwen2.5-Coder\xspace}
\newcommand{\deepseekcoder}{DeepSeekCoder\xspace}
\newcommand{\starcoder}{StarCoder2\xspace}
\newcommand{\qwencodersmall}{Qwen2.5-Coder-1.5B\xspace}
\newcommand{\qwencoderbig}{Qwen2.5-Coder-7B\xspace}
\newcommand{\deepseeksmall}{DeepSeekCoder-1.3B\xspace}
\newcommand{\deepseekbig}{DeepSeekCoder-6.7B\xspace}

\newcommand{\starcoderbig}{StarCoder2-7B\xspace}
\newcommand{\codereval}{CoderEval\xspace}

\newcommand{\codellama}{CodeLlama-7B\xspace}

\definecolor{myyellow}{HTML}{FFF2CC}

\newcounter{finding}
\newcommand{\finding}[1]{\refstepcounter{finding}
 	\vspace{1mm}
	\begin{mdframed}[linecolor=gray!25,roundcorner=12pt,backgroundcolor=myyellow!30,linewidth=3pt,innerleftmargin=2pt, leftmargin=0cm,rightmargin=0cm,topline=false,bottomline=false,rightline = false]
		\textbf{Finding \arabic{finding}:} #1
	\end{mdframed}
	\vspace{1mm}
}

\newcommand{\boxmargin}{1mm}
\tcbuselibrary{most, skins, breakable, theorems}
\newtcolorbox{myboxa}[2][]{
    colback=gray!10!white,
    colframe=black, enhanced,
    attach boxed title to top left={yshift=-2mm,xshift=5mm},
    title=#2,#1
}
\newtcolorbox{myboxb}[2][]{
    boxsep=3pt,
    left = \boxmargin, right = \boxmargin, top = \boxmargin, bottom = \boxmargin,
    title={#2},#1
}
\newtcolorbox{myboxc}{
    colback=gray!15!white,
    arc = 0pt, outer arc = 0pt,
    boxsep=0pt, left = 3pt, right = 0pt, top = 0pt, bottom = 0pt, 
    leftrule=3pt, bottomrule=0pt,toprule=0pt, rightrule=0pt,
    left = \boxmargin, right = \boxmargin, top = \boxmargin, bottom = \boxmargin
}
\newtcolorbox{myboxd}{
    colback=gray!10,
    colframe=black,
    width=\columnwidth,
    arc=1mm, auto outer arc,
    boxrule=0.5pt,
}

\usepackage{tikz}

\newcommand{\figmargin}{\vspace{-2pt}}

\begin{document}
\pagestyle{empty}

\title{A Preliminary Study on the Robustness of Code Generation by Large Language Models}

\newcommand\corrauthorfootnote[1]{%
  \begingroup
  \renewcommand\thefootnote{}\footnote{\textsuperscript{*}#1}%
  \addtocounter{footnote}{-1}%
  \endgroup
}

\author{Zike Li}
\affiliation{ 
  \institution{School of Software Engineering\\ Sun Yat-sen University}
  \city{Zhuhai}
  \country{China}}
\email{lizk8@mail2.sysu.edu.cn}
 
\author{Mingwei Liu$^{\ast}$}
\affiliation{ 
  \institution{School of Software Engineering\\ Sun Yat-sen University}
  \city{Zhuhai}
  \country{China}}
\email{liumw26@mail.sysu.edu.cn}

\author{Anji Li}
\affiliation{ 
  \institution{School of Software Engineering\\ Sun Yat-sen University}
  \city{Zhuhai}
  \country{China}}
\email{lianj8@mail2.sysu.edu.cn}

\author{Kaifeng He}
\affiliation{ 
  \institution{School of Software Engineering\\ Sun Yat-sen University}
  \city{Zhuhai}
  \country{China}}
\email{hekaifeng70@gmail.com}

\author{Yanlin Wang}
\affiliation{ 
  \institution{School of Software Engineering\\ Sun Yat-sen University}
  \city{Zhuhai}
  \country{China}}
\email{wangylin36@mail.sysu.edu.cn}

\author{Xin Peng}
\affiliation{ 
  \institution{School of Computer Science\\ Fudan University}
  \city{Shanghai}
  \country{China}}
\email{pengxin@fudan.edu.cn}

\author{Zibin Zheng}
\affiliation{ 
  \institution{School of Software Engineering\\ Sun Yat-sen University}
  \city{Zhuhai}
  \country{China}}
\email{zhzibin@mail.sysu.edu.cn}

\begin{abstract}

Robustness is a critical factor for reliable code generation by large language models, yet most evaluations focus on correctness and overlook key issues such as missing input validation and inadequate error handling. In this work, we present the first empirical study of LLM-generated code robustness using the CoderEval benchmark. Evaluating four state-of-the-art code LLMs, we find that 35.2\% of their outputs are less robust than human-written code, with over 90\% of deficiencies caused by missing conditional checks—70\% of which occur in the first line. Interestingly, in 63\% of cases where a conditional statement is needed but absent, the “if” token still ranks among the top three predictions, suggesting implicit recognition of control flow.

To address these issues, we propose RobGen, a model-agnostic framework that improves robustness without retraining. RobGen combines a line-level intervention checker, which decides whether to adjust logits for each generated line, with token-level conditional logit adjustments to promote essential control structures. Experiments show that RobGen reduces the proportion of less robust code by 10\%, achieves the highest average Pass@1 (43.57), and adds minimal overhead (+33.4\%). As a lightweight and adaptable solution, RobGen effectively enhances the reliability of LLM-generated code across diverse tasks.
\end{abstract}


\maketitle

\section{INTRODUCTION}

Automatic code generation, which involves synthesizing code snippets that fulfill specified requirements, has become a crucial aspect of modern software engineering~\cite{openai2021codex, zhang2023draft, radford2018improving, dong2023self, jiang2023selfevolve, zan2023large}. The emergence of large language models for code (Code LLMs), such as CodeLlama~\cite{roziere2023code}, StarCoder~\cite{li2023starcoder}, DeekSeekCoder~\cite{guo2024deepseek} and Qwen2.5-Coder~\cite{qwen2.5}, has significantly advanced the capabilities of automated code generation~\cite{chen2021evaluating, nijkamp2023codegen, rozière2024code, li2023starcoder, guo2024deepseekcoder}. While prior research has primarily focused on improving the correctness of generated code~\cite{ugare2024improving,su2024distilled,islam2024map}, robustness remains relatively underexplored. Ensuring robustness is essential for handling edge cases, invalid inputs, and unexpected execution scenarios.

 Although many studies measure LLM-generated code correctness using pass@k, such metrics fall short of capturing robustness. Even if a model passes all tests, it may still lack necessary robustness checks~\cite{liu2024your}. Conversely, a model might fail some tests while its core logic is correct, but it may miss critical validations such as handling empty inputs (see Figure\ref{example}(a)), the code fails to pass the task tests due to missing input parameter checks for ``str'' and ``searchStrArray''. These issues underscore that unit test pass rates alone do not fully reflect code robustness. This gap motivates our study, where we develop new robustness metrics and conduct an in-depth empirical investigation into the robustness deficiencies of LLM-generated code.

 
 Recently, Liu et al. \cite{LiuYue2024Quality} conducted a systematic analysis of ChatGPT-generated code, evaluating its correctness while also identifying potential quality issues. Zhong et al. \cite{Zhong_Wang_2024} introduced the RobustAPI dataset to assess the reliability and robustness of LLM-generated code. However, these studies did not deeply explore the robustness of model-generated code. To improve code robustness, Zhang et al. \cite{zhang2024seekerenhancingexceptionhandling} proposed Seeker, a multi-agent framework to generate high-quality exception-handling code. However, the use of a multi-agent system requires multiple LLM invocations, leading to significant time overhead.


\textbf{Empirical Study.}This work presents the first comprehensive study on the robustness of LLM-generated code using CoderEval, a benchmark for complex repository-level code generation. Unlike prior approaches that primarily rely on unit tests, our methodological contribution lies in leveraging LLM as evaluator to directly assess code properties, complemented by comparisons with human-written ground truth for a more comprehensive and fine-grained robustness evaluation. The study analyzes four Code LLMs, investigating (1) the robustness gap between generated and human-written code, (2) prevalent patterns of robustness issues, (3) the specific locations where these issues arise, and (4) whether LLMs inherently recognize the need for robustness checks.

Our empirical study reveals that on average 37.8\% of LLM-generated code is less robust than human-written code, with significant room for improvement. We identify nine distinct patterns of robustness issues, \textbf{90\%} of which are related to missing conditional checks. Further analysis shows that \textbf{70\%} of these issues occur in the \textbf{first line} of the generated code. Additionally, in these patterns, the \textbf{"if" token ranks} third or higher in \textbf{69\%} of the cases where conditionals should be generated but are missing.

\textbf{Plug-in Framework.} 
Building on our findings, we introduce \app, a plug-and-play framework for enhancing the robustness of LLM-generated code without retraining. \app consists of two model-agnostic techniques: line-level intervention checker and token-level conditional logit adjustment. 
During code generation, the line-level intervention checker inspects each line to decide if a logit adjustment should be applied. token-level conditional logit adjustment operates during decoding, dynamically adjusting token logits to encourage the generation of essential control structures. As a lightweight and adaptable solution, \app effectively improves code reliability across different models and tasks while fully leveraging the model’s inherent capabilities. We leverage LLM as evaluator to measure the robustness difference between generated and reference code. Experimental results across five models show that \app reduces the proportion of less robust code by 10\% and achieves the highest average Pass@1 (43.57), enhancing code generation robustness with minimal time overhead (+33.4\%), thereby demonstrating its effectiveness.


We summarize the main contributions of this paper as follows.
\begin{itemize}[left=1mm, labelsep=1em,itemsep=2pt,topsep=0pt,parsep=0pt]
    \item Conducting the first in-depth empirical analysis of robustness in LLM-generated code.  
    \item Leveraging LLM to assess robustness, focusing on edge cases, invalid inputs, and error handling.  
    \item Introducing \app, a plug-in framework with two lightweight, model-agnostic techniques that refine control structures during and after generation without retraining.  
\end{itemize}


\section{EMPIRICAL STUDY}
\label{sec:empircal}
To assess the robustness of LLM-generated code, we conduct an empirical study with the following research questions (RQs):
\begin{itemize}[left=1mm, labelsep=1em,itemsep=2pt,topsep=0pt,parsep=0pt]
    \item \textbf{RQ1: How robust is the code generated by LLMs compared to human-written code?} We employ an LLM-based evaluator to assess the robustness of LLM-generated code by comparing it to human-written code, and further analyze statistical metrics to capture concrete aspects of robustness.
        
    \item \textbf{RQ2: What are the common patterns of robustness issues in LLM-generated code?} We analyze cases where LLM-generated code is less robust than the reference, identifying recurring issues like missing null checks to reveal weaknesses and inform improvement strategies.

    \item \textbf{RQ3: Where do robustness issues tend to occur in LLM-generated code?} We investigate the distribution of robustness issues within the generated code at the line level.

    \item \textbf{RQ4: Do LLMs recognize the need for condition statements to improve robustness?} We analyze token probability distributions at positions where an if-statement is expected to determine whether LLMs inherently recognize the necessity of condition statements to enhance code robustness.
\end{itemize}

\subsection{Experiment Setup}
\label{sec:setting}
This section outlines the empirical study settings, including model selection, task selection, and implementation details. 

\subsubsection{Model Selection}
We selected four mainstream and representative open-source Code LLMs that have demonstrated strong performance in code generation: \qwencodersmall~ and \qwencoderbig~ (September 2024) \cite{hui2024qwen25codertechnicalreport}, as well as \deepseeksmall~ and \deepseekbig~ (January 2024) \cite{guo2024deepseekcoder}. These models are instruction-tuned versions of their base models, enhancing their ability to follow instructions and improving their performance in code generation tasks~\cite{qwen2.5,deepseek-llm}.

All models were sourced from official sources and used according to their guidelines. Due to resource constraints, we focus on models with fewer than 7B parameters. Given the limitations of smaller models, ensuring the robustness of their generated code becomes even more critical. Using open-source models also allows for detailed analysis of token probability distributions, offering valuable insights into model behavior.

\subsubsection{Task Selection}
\label{sec:setting:tasks}

To evaluate LLM-generated code robustness, we selected CoderEval~\cite{yu2024codereval} as our benchmark. Unlike OpenAI’s HumanEval~\cite{chen2021codex}, which focuses on function-level generation without context, CoderEval offers a more realistic setting with repository-level context, increasing complexity by requiring models to handle dependencies and produce robust code. CoderEval includes 230 Java tasks from real-world open-source projects, each with a docstring, function signature, ground truth implementation, and unit tests. Java’s widespread use in enterprise applications~\cite{ouh2023chatgpt}, static type system, and strict exception handling~\cite{st2024toward} make it ideal for assessing robustness in LLMs.

\subsubsection{Implementation Details}
We configure all Code LLMs according to the official guides, using the default settings. To ensure a fair comparison, the maximum token limit is set to 300. During the inference stage, we employ a greedy sampling strategy~\cite{Goodfellow2016} to avoid randomness across the four Code LLMs. All experiments are conducted on a machine equipped with 512 GB of RAM and an Nvidia A800 GPU with 80 GB of memory.

\begin{figure*}[!t]
	\centering
\includegraphics[width=0.9\textwidth]{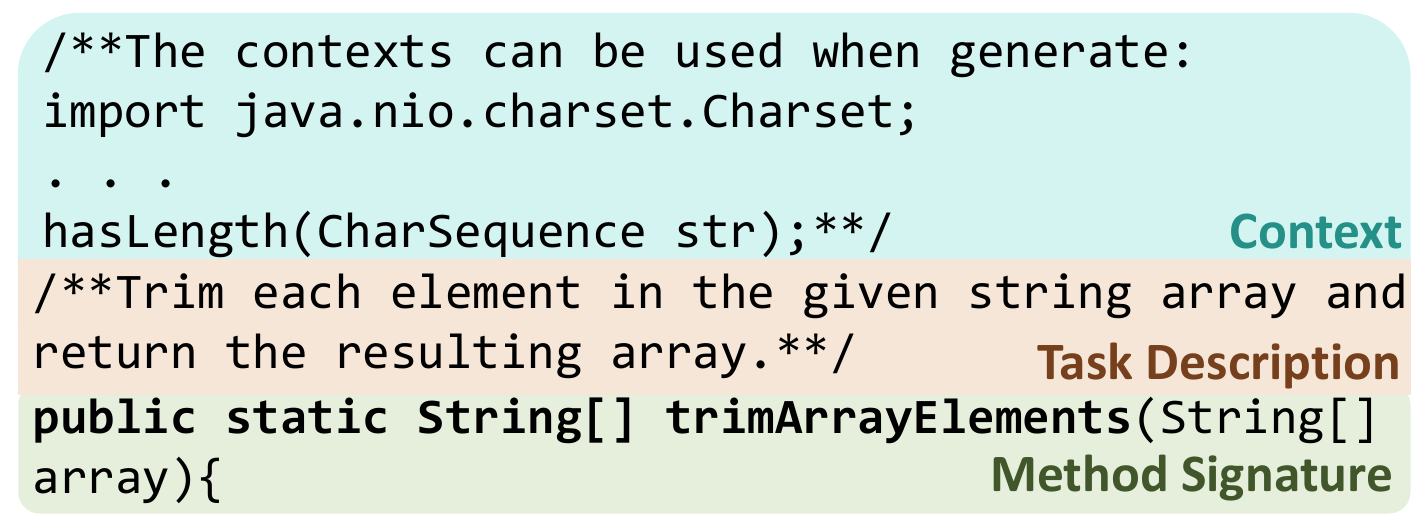}
	\caption{Code Generation Prompt Used in Experiments}
        \label{temple}
\end{figure*}

Figure~\ref{temple} shows an example of the prompts used in our experiments, based on previous work~\cite{WhentoStop2024}. For each task, we combine the task description, context, and method signature as input and ask the LLMs to generate the full method. In some cases, models generate extra code beyond the given method signature, as noted in prior studies~\cite{WhentoStop2024}. To improve performance, we use a rule-based matching approach to filter out unnecessary code.

\subsection{RQ1: Robustness of LLM-Generated Vs. Human-Written Code}
\label{sec:rq1}
To answer this RQ, we compare the robustness of LLM-generated code with the ground truth implementations, which are human-written code from real-world projects. For the robustness evaluation, we employ both a LLM as an automatic evaluator and complementary statistical metrics to provide quantitative evidence.

\subsubsection{Analysis Target Selection}
As described in Section~\ref{sec:setting}, we use 230 Java coding tasks from CoderEval and generate one code snippet per task using different models with greedy decoding. Table~\ref{tab:Pass1Compile1} presents the Compile@1 (i.e., the number of generated code snippets that successfully compile without errors) and Pass@1 (i.e., the number of snippets that pass all test cases), along with the corresponding counts for each model. We filter out LLM-generated code that fail to compile as such code is often incomplete or contains undefined methods and variables.

\subsubsection{LLM-Based Automatic Evaluator}
\label{sec:rq1:judger}
\begin{table*}[]
\centering
\caption{Compile@1 and Pass@1 for Studied LLMs}
\label{tab:Pass1Compile1}

\begin{tabular}{|c|c|c|}
\hline

Model           & \textit{Compile@1} & \textit{Pass@1} \\ \hline
\deepseeksmall  & 0.67 (153)          & 0.34 (79)        \\ \hline
\deepseekbig    & 0.75 (173)          & 0.46 (105)       \\ \hline
\qwencodersmall & 0.67 (154)          & 0.40 (91)        \\ \hline
\qwencoderbig   & 0.75 (172)          & 0.49 (112)       \\ \hline
\end{tabular}%

\end{table*}
To evaluate code robustness at scale, we adopt a LLM as an automatic evaluator. Using LLMs as “judges” has become common in tasks such as code quality assessment, summarization evaluation, and natural language generation benchmarking~\cite{BUSKER2025101988,info15020099}, due to their strong analytical and contextual reasoning abilities. Compared with manual evaluation, this approach is cost-efficient, reproducible, and avoids inter-rater variability~\cite{cheng2023gpt4gooddataanalyst}.

\begin{figure*}[!t]
	\centering
\includegraphics[width=0.9\textwidth]{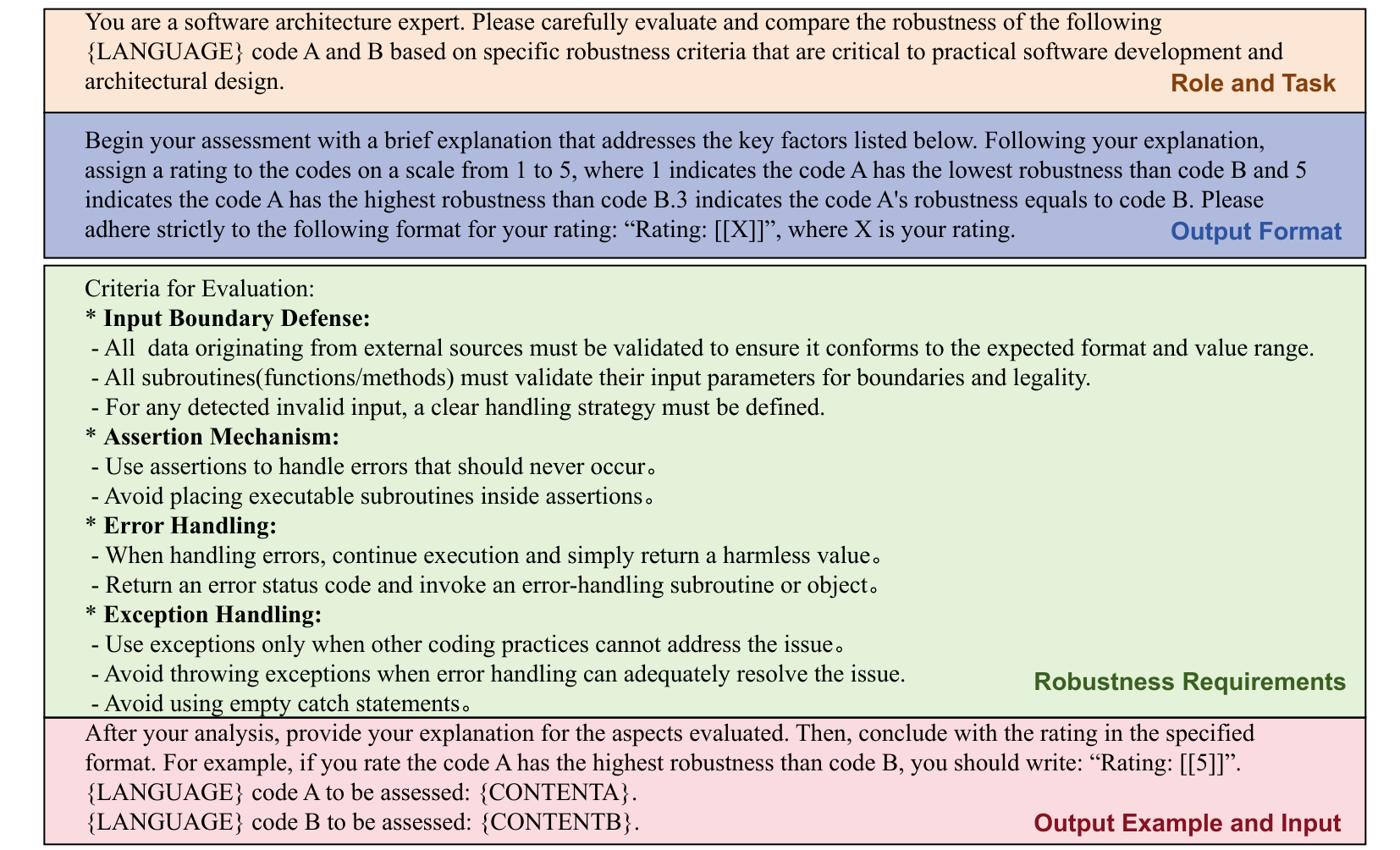}
	\caption{Prompt for Evaluating Robustness}
        \label{robust_prompt}
\end{figure*}

We employ GPT-4o as the evaluator to compare two code snippets (LLM-generated vs. human-written). Fig.~\ref{robust_prompt} illustrates the prompt template for assessing a code pair in a specified language. The evaluation follows \emph{defensive programming principles}~\cite{denfensiveprograming}, which emphasize boundary checking, exception handling, and input validation—core practices that directly relate to robustness. Guided by these criteria, the prompt instructs the LLM to assign a comparative robustness score. To avoid position bias, Code A and Code B are randomly assigned, hiding their origin from the evaluator. Scores are on a 1–5 scale: $>3$ indicates Code A is more robust, $<3$ less robust, and 3 denotes comparable robustness.

To reduce variance in generative outputs, each comparison is repeated three times and averaged. We validate this procedure by sampling 100 code pairs, performing manual robustness comparisons, and computing Cohen’s Kappa~\cite{CohenKappa}, yielding 0.81, indicating near-perfect agreement. An alternative \emph{independent scoring} approach (absolute scores 0–10 per snippet) gives a Kappa of 0.62, showing lower alignment with human judgment.
All prompts, scoring instructions, and annotations are included in the released package~\cite{Robgen}. The evaluation uses GPT-4o via API calls, and details of hyperparameters (temperature, max tokens, etc.) are provided in the release. GPT-4o was chosen for its strong reasoning and code understanding capabilities, enabling scalable and reliable robustness assessment without model fine-tuning~\cite{cheng2023gpt4gooddataanalyst}.

\subsubsection{Statistical Metrics Analysis}
\label{sec:rq1:metircs}
Beyond LLM judgments, we employ statistical metrics to capture concrete aspects of robustness, specifically examining control expressions and exception handling as indicators of defensive programming practices.


\parabf{Control Expressions Analysis.}
To assess the structure and complexity of control logic, we analyze \textbf{atomic Boolean expressions}, defined as the smallest evaluable Boolean conditions in control statements (e.g., \texttt{if}, \texttt{for}, \texttt{while}). Examples include simple checks like \texttt{x > 5} or \texttt{isValid()}. Using abstract syntax trees (ASTs) extracted via Tree-sitter~\cite{treesiter}, we identify all atomic Boolean expressions in both generated and reference code.

We define the following metric:

\textbf{Average Atomic Boolean Expressions (AvgABE):} measures the average number of atomic Boolean expressions per snippet:
\begin{equation}
AvgABE = \frac{1}{N} \sum_{i=1}^{N} |U_i|
\end{equation}
where \(U_i\) is the set of atomic Boolean expressions in snippet \(i\) and \(N\) is the total number of snippets. A higher AvgABE indicates more explicit conditions, suggesting stronger robustness through enhanced boundary checks and input validation.

\parabf{Exception Handling Analysis.}  
Exception handling is a critical component of robust software. We define:

\textbf{Exception Handling Adoption Rate (EHAR):} the proportion of code snippets that include \texttt{try-catch} blocks:
\begin{equation}
EHAR = \frac{\text{\# of snippets with try-catch}}{\text{Total \# of snippets}}
\end{equation}
A higher EHAR reflects greater reliance on exception handling, which can help capture runtime errors but may also indicate heavier dependence on reactive rather than preventive mechanisms.

\begin{table}[]
    \centering
    \caption{Comparison of Robustness Metrics Across Models: ``DSC'' for DeepSeekCoder and ``QWC'' for Qwen2.5-Coder, ``Gen'' means Generated Code, ``GT'' means Ground True Code}
    \label{RQ1Metric}
    {
    \begin{tabular}{|c|cc|cc|}
    \hline
    \multirow{2}{*}{\textbf{Model}} & \multicolumn{2}{c|}{\textbf{EHAR}}                 & \multicolumn{2}{c|}{\textbf{AvgABE}}               \\ \cline{2-5} 
                                    & \multicolumn{1}{c|}{Gen}           & GT            & \multicolumn{1}{c|}{Gen}           & GT            \\ \hline
    DSC-1.3B                        & \multicolumn{1}{c|}{0.02}          & \textbf{0.03} & \multicolumn{1}{c|}{1.39}          & 2.04          \\ \hline
    DSC-6.7B                        & \multicolumn{1}{c|}{\textbf{0.03}} & 0.02          & \multicolumn{1}{c|}{1.46}          & 2.10          \\ \hline
    QWC-1.5B                        & \multicolumn{1}{c|}{0.01}          & \textbf{0.03} & \multicolumn{1}{c|}{1.62}          & \textbf{2.12} \\ \hline
    QWC-7B                          & \multicolumn{1}{c|}{\textbf{0.03}} & \textbf{0.03} & \multicolumn{1}{c|}{\textbf{1.83}} & 2.02          \\ \hline
    \end{tabular}
    }

\end{table}
\begin{figure*}[!t]
	\centering
\includegraphics[width=0.9\textwidth]{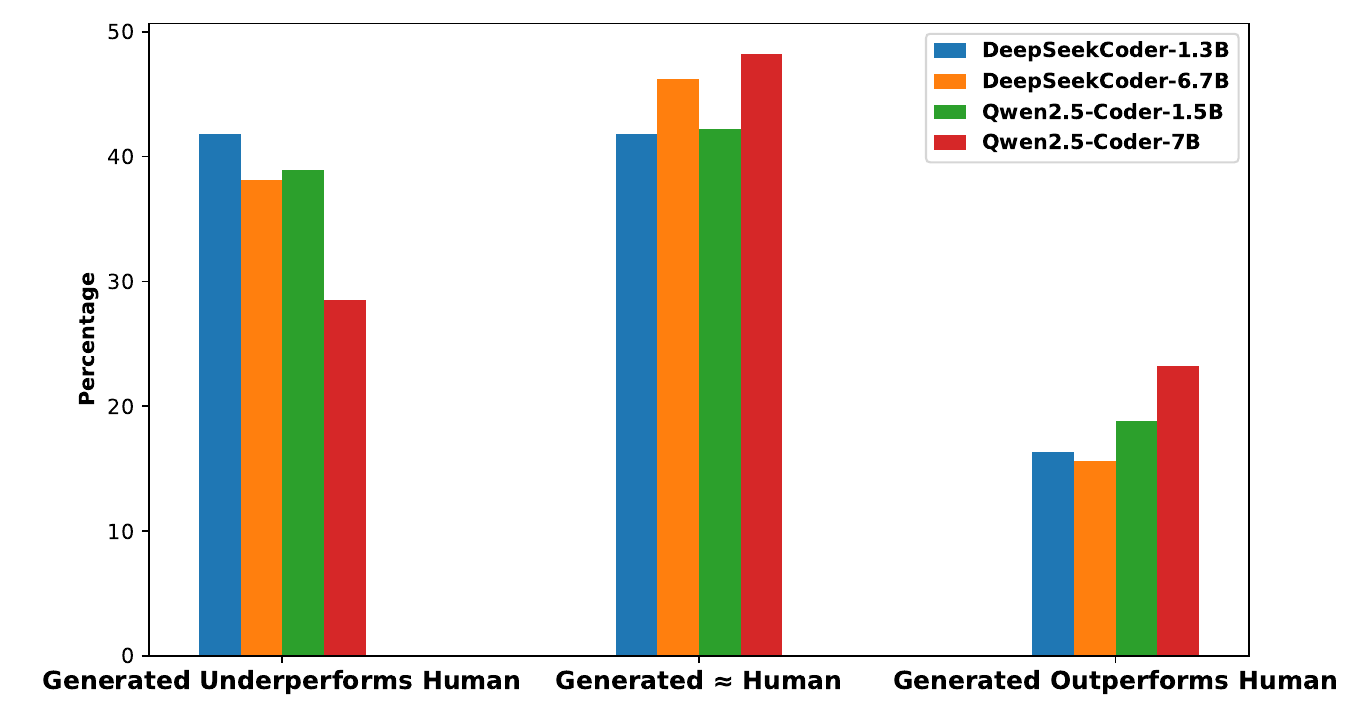}
	\caption{GPT-4o as a Judger: Results. ``Generated Underperforms Human'' indicates that the LLM judges human written code to be more robust than LLM generated code, the same applies to other labels.}
        \label{RQ1Judger}
\end{figure*}
\subsubsection{Results}
For the selected LLM-generated code, We examined the distribution of LLM Evaluator scores measuring the relative robustness of generated code compared to human written counterparts, and computed control expression and exception handling metrics relative to the reference code. The results are presented in Table \ref{RQ1Metric} and Figure \ref{RQ1Judger}.

\parabf{LLM-Based Automatic Evaluator Results Analysis.}
Figure \ref{RQ1Judger} shows the distribution of tasks where generated code exhibits lower, equal, or better robustness compared to reference code. On average, 35.2\% of generated code is less robust, 45.1\% is equally robust, and 19.5\% is more robust. The gap between cases where human code outperforms and underperforms generated code highlights substantial room for improvement.


For \qwencodersmall, 39.0\% of compiled code is less robust than the reference code, while for \qwencoderbig, this proportion decreases to 28.5\%. This suggests that increasing the parameter size of the \qwencoder models leads to improved robustness in generated code. These results suggest that, overall, larger models tend to generate more robust code. When comparing models of similar parameter sizes, \qwencoder exhibits superior robustness in code generation compared to \deepseekcoder. Specifically, for \qwencodersmall, the proportion of code with ``Human poorer than Generated'' is 18.8\%. In contrast, for \deepseeksmall, the proportion is 16.3\%, respectively, indicating slightly weaker robustness. 
A similar trend is observed in the \deepseekcoder series: for \deepseekcoder 1.3B and 6.7B, the proportions of generated code with lower robustness compared to the reference code are approximately 41.8\% and 31.2\%, respectively. 

\parabf{Control Expression Analysis.}
We performed a Chi-square test~\cite{Pearson1900} to assess the association between the number of atomic expressions and code robustness. The result revealed a statistically significant association between the number of atomic expressions and the robustness of generated code (p << 0.05). The corresponding Cramér’s V of 0.47 indicates a moderate-to-strong relationship, suggesting that a higher number of atomic expressions is meaningfully linked to increased robustness in generated code.

The AvgABE metric quantifies the average number of atomic Boolean expressions per code snippet. As indicated in Table \ref{RQ1Metric}, for the ground truth code, the AvgABE values across different subsets of model-generated code consistently remain around 2.0, suggesting that each ground truth snippet contains approximately two atomic Boolean expressions on average. In contrast, model-generated code tends to exhibit lower AvgABE values. Notably, \deepseeksmall demonstrates the lowest AvgABE value of 1.39. Among the evaluated models, \qwencoderbig performs relatively well, achieving an AvgABE of 1.83. 

Additionally, we observe systematic differences across model series: the \qwencoder series generally produces code with higher AvgABE values compared to the \deepseekcoder Coder series. For example, \qwencodersmall generates code with an AvgABE value exceeding that of \deepseekbig.

\parabf{Exception Handling Analysis.} 
The EHAR metric quantifies the proportion of code with \texttt{try-catch} blocks. As shown in Table \ref{RQ1Metric}, the EHAR values for ground truth and model-generated code remain consistently around 0.03, with minimal variation. It indicates that about 3\% of the ground truth code includes explicit exception handling, reflecting the low prevalence of \texttt{try-catch}. This is reasonable, as best practices discourage overusing \texttt{try-catch}, with foreseeable exceptions handled using guard clauses.
Similarly, the EHAR values for model-generated code are low. \qwencodersmall has a value of 0.01 (1\%), while \qwencoderbig rises to 0.03, matching the prevalence in human-written code.


\finding{LLM evaluator results show that LLM-generated code is on average 81.5\% which robustness is less than or equal to human-written code. Robustness metrics further indicate that control expressions warrant more attention than exception handling, as they occur more frequently in real-world code and exhibit a larger performance gap.}


\subsection{RQ2: Patterns of Robustness Issues in LLM-Generated Code}
\label{sec:rq2}
We investigate common patterns of robustness issues in LLM-generated code by analyzing cases where the generated code is less robust than the reference implementation. 

\subsubsection{Data Selection}  
The dataset used for this analysis originates from RQ1, consisting of 920 generated code snippets from four models (230 snippets per model). 
Given the high cost of full manual analysis, we use LLM evaluator as an initial filtering step to retain only code snippets exhibiting lower robustness than the reference implementation, resulting in a refined dataset with 364 code snippets for detailed manual inspection.

Further, to maintain the continuity and consistency of the analysis, we follow the task selection strategy employed in RQ1~\ref{sec:rq1} and exclude all code snippets that cannot be successfully compiled. As a result, 220 code snippets are kept for further analysis.


\subsubsection{Analysis Process}  
For each code snippet, we analyze whether robustness issues exist in the code snippet and what category can it be categorized into. For example, some robustness issues are arise from flaws in Boolean logic expressions affecting control flow, such as missing conditions, incorrect conditions, or inadequate input validation. Using an open coding approach~\cite{strauss1990basics}, we iteratively categorize these issues, refining the categories as needed. A single snippet may exhibit multiple issues and receive multiple labels. Two authors independently annotate the data, resolving disagreements through discussion and majority consensus.The computed kappa agreement score is 0.71.

To ensure accurate assessment, we consider additional contextual information, including reference code, surrounding class context, compilation feedback, and evaluation outputs of the LLM evaluator.
For complex cases where LLM-generated code implements multiple functionalities, each part is individually compared against the reference implementation. The comparison focuses on semantic equivalence rather than surface-level differences; for instance, expressions like ``\texttt{child == null}'' and ``\texttt{childNode == null}'' are considered equivalent if they serve the same logical purpose.

\begin{figure*}[!t]
        
	\centering
\includegraphics[width=0.9\textwidth]{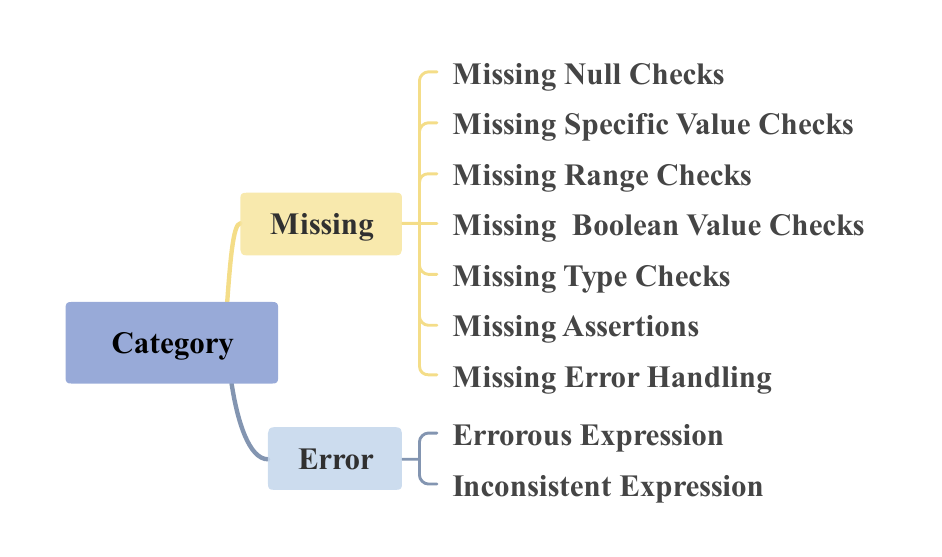}
	\caption{Category of Poor Robustness Pattern}
        \label{category}
\end{figure*}
\subsubsection{Results}
\begin{figure*}[!t]
        
	\centering
        \includegraphics[width=0.9\textwidth]{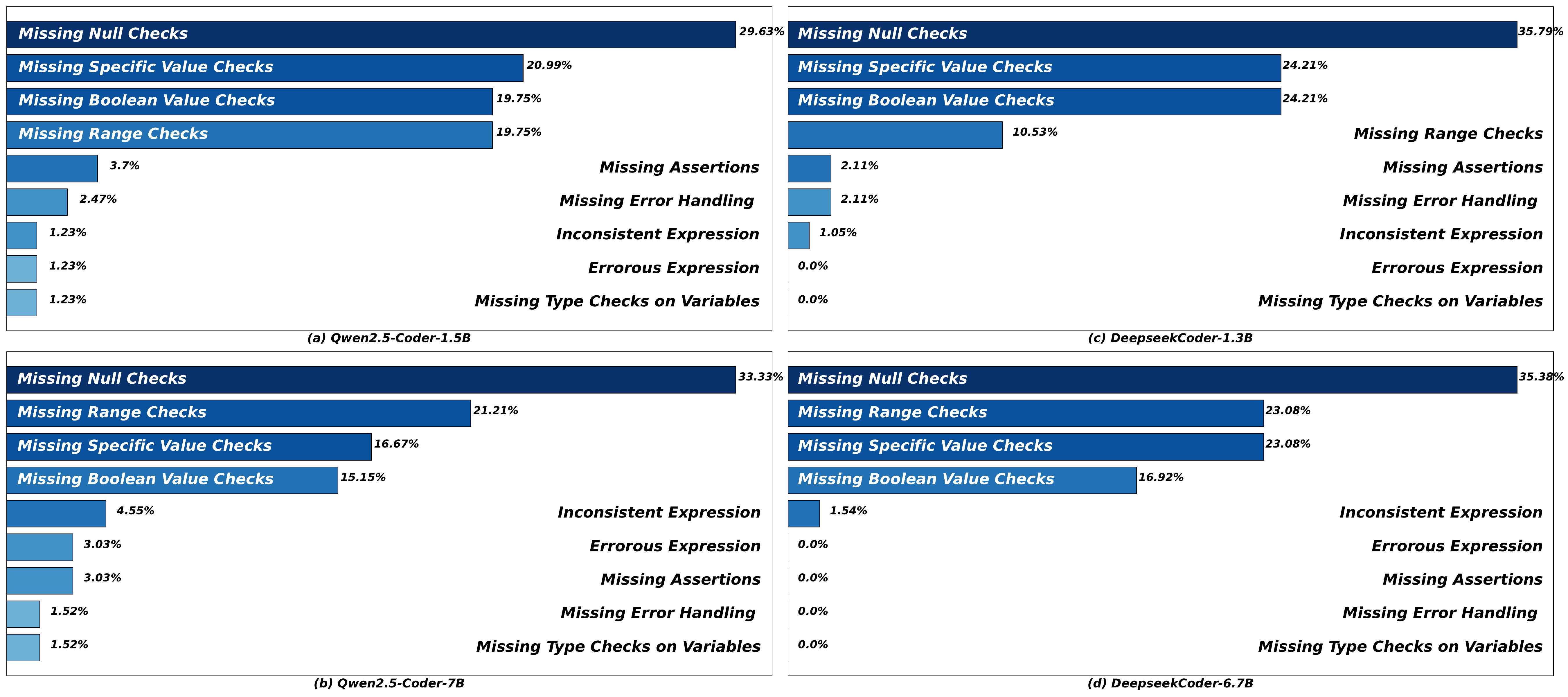}
	\caption{Distribution of Robustness Issue Patterns}
        \label{classification}
\end{figure*}

Based on our observations, we divided the situations where code lacks robustness into two main categories: Missing  and Errors, as shown in Figure \ref{category}.
\begin{itemize}

\item \textbf{Missing:} This indicates that the model has not recognized the need for an additional check to improve the code's robustness at this point.

\item \textbf{Error:} An error occurs when the model identifies the necessity of generating a robustness check but produces an incorrect condition, thereby compromising the functionality or even the correct execution of the code.
\end{itemize}






For the Missing category, we further divided it into seven subcategories:
\begin{itemize}[left=1mm, labelsep=1em,itemsep=2pt,topsep=0pt,parsep=0pt]

\item \textbf{Missing Null Checks:} A well-written piece of code should check input variables for null pointers as much as possible to prevent operations on null objects that could crash the system, especially in complex software systems where receiving a null object is common in certain modules.

\item \textbf{Missing Specific Value Checks:} A typical example of this is checking whether an array’s length is zero. In engineering practice, checking the length of an array effectively prevents out-of-bounds errors. Generally, checking for specific values in the code is to detect special situations and handle them accordingly.

\item \textbf{Missing Range Checks:} Range checks are usually used to ensure that the value of some variables or indices do not exceed a certain boundary to avoid errors and exceptions.

\item \textbf{Missing Boolean Value Checks:} Some project code may contain global Boolean variables that indicate the current state of the project. Checking these global Boolean variables helps the program make decisions based on the current state of the project.

\item \textbf{Missing Type Checks:} Type checking directly prevents incorrect operations on types or helps the program execute different actions based on the variable's type.

\item \textbf{Missing Assertions:} In some code, assertions are used for defensive programming. The role of assertions is similar to that of If statements, and they are often found frequently in unit test code. We observed that the model tends to use If statements for defense when generating non-test unit code, rather than using assertions.

\item \textbf{Missing Error Handling:} The \texttt{try-catch} structure is essential for handling exceptions in I/O operations, file handling, and directory creation, preventing crashes and ensuring proper error management. However, we observed that in critical scenarios requiring \texttt{try-catch} for stability, the model sometimes fails to generate it correctly.

\end{itemize}

For the Error category, we further divided it into two subcategories:

\begin{itemize}[left=1mm, labelsep=1em,itemsep=2pt,topsep=0pt,parsep=0pt]
\item \textbf{Errorous Expression:} CodeLLM may generate code with undefined methods or variables, causing compilation failures. We also observed this issue in boolean atomic expressions, affecting code correctness and reliability.

\item \textbf{Inconsistent Expression:} Expecting the model's generated code to be identical to standard code is unrealistic. The inconsistency in this context refers to variations in boundary condition judgments within atomic boolean expressions. For example, expressions like \texttt{A <= B} vs. \texttt{A < B} or \texttt{len < head} vs. \texttt{len < size} illustrate such inconsistencies.

\end{itemize}

We analyzed the distribution of the previously identified patterns, and the results are shown in Figure \ref{classification}. As illustrated, around \textbf{90\%} of the occurrences across the nine patterns are due to \textbf{Missing Null Checks}, \textbf{Missing Specific Value Checks}, \textbf{Missing Range Checks}, and \textbf{Missing Boolean Value Checks}, all of which involve missing conditional checks. Among these, \textbf{Missing Null Checks} is the most common pattern.

Interestingly, for models with fewer parameters, such as \deepseeksmall, \textbf{Missing Specific Value Checks} as the second most frequently observed pattern. In contrast, for larger models, such as \deepseekbig, \textbf{Missing Range Checks} becomes the second most common pattern. Other patterns within the Missing category, such as \textbf{Missing Error Handling}, as well as those in the Error category, occur at significantly lower frequencies.

\finding{We identified nine distinct robustness patterns. Statistical analysis of their frequencies revealed that over 90\% of these issues are related to missing conditional checks, with Missing Null Checks being the most prevalent.}

\subsection{RQ3: Distribution of Robustness Issues}
\label{sec:rq3}
In this RQ, we analyze the line-level distribution of robustness issues in generated code.

\subsubsection{Design}
Building on RQ2, where we identified robustness issues in LLM-generated code along with their corresponding patterns, we further analyze the specific locations where these issues manifest. For code containing multiple robustness issues, we record only the first occurrence, as LLMs generate code token by token, meaning earlier tokens influence subsequent ones. Thus, capturing the initial occurrence provides insight into the root cause of robustness deficiencies.

For issues categorized under the Error patterns, the recorded occurrence corresponds to the exact line containing the erroneous construct, such as an incorrect if condition. For issues categorized under the Missing patterns, determining the precise location requires structural alignment with the reference implementation. In such cases, annotators manually examine the reference code to identify where the missing element should have been placed. The designated location is chosen to ensure minimal deviation in control flow between the generated code and the reference implementation. For example, if the generated code omits an essential input validation check and lacks any guard conditions, and the reference code places this check at the beginning, the robustness issue is recorded at the first line. This approach maintains consistency and ensures an objective comparison.

\subsubsection{Results}
\begin{figure*}[!t]
	\centering
\includegraphics[width=0.9\textwidth]{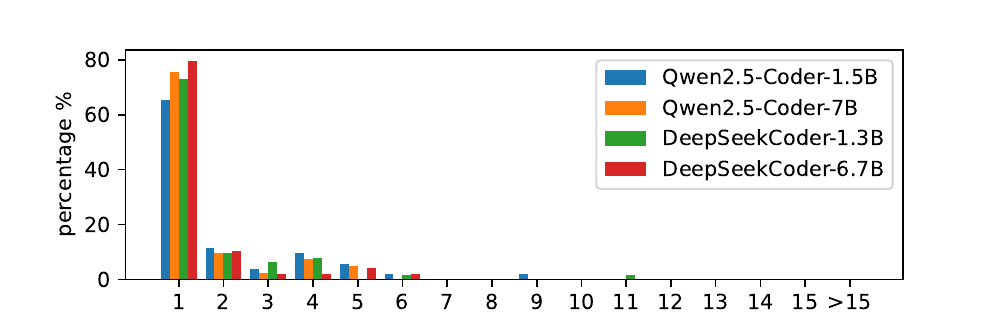}
	\caption{Line-level Distribution of Robustness Issues}
        \label{distribution}
\end{figure*}
 
Figure \ref{distribution} illustrates the positional distribution of robustness issues in the generated code for the four evaluated LLMs. Most issues (70\%) occur in the first line, primarily due to missing robustness checks, such as input validation. This aligns with expectations, as critical checks like null pointer validation are typically placed at the beginning of a function or method.   

A detailed analysis shows that \deepseekbig exhibits the highest proportion (79.6\%) of robustness issues in the first line, while \qwencoderbig has a lower rate (75.6\%). Model size also affects the likelihood of first-line robustness issues. As model size increases, the proportion of missing checks increases. For instance, \deepseeksmall has a 73.0\% issue rate, which grows to 79.6\% in \deepseekbig. Similarly, \qwencodersmall has a 65.4\% rate, rising to 75.6\% in \qwencoderbig.  Additionally, \deepseekcoder models generate a higher proportion of first-line robustness issues compared to \qwencoder models, suggesting that differences in training data or model architecture impact robustness of generated code.

\finding{Most robustness issues occur in the first line of the generated code, primarily due to missing robustness checks.}

\subsection{RQ4: Condition Generation Potential}
\label{sec:rq4}
We analyze token probabilities at expected ``if'' statement positions to assess whether LLMs inherently recognize the need for control structures essential for robust code, even if they sometimes omit them during greedy decoding.

\subsubsection{Design}
As indicated in RQ2, a large fraction (90\%) of robustness issues in generated code stem from missing condition statements. In earlier experiments, all LLM-generated code was produced using greedy decoding, which always selects the token with the highest logit. To assess the model’s latent understanding, we analyze the ranking distribution of the ``if'' token in the model’s logit output at positions where an if-statement is expected. Our assumption is that if the ``if'' token is ranked highly at these positions, the model possesses the inherent capability to generate a complete condition statement, thereby mitigating robustness issues.

Building on the findings from RQ3, we first identify code lines where missing condition checks lead to robustness issues. For each identified line, we determine the expected insertion point—marked by the first non-whitespace token—and exclude cases where an ``if'' statement already exists. This filtering yields 194 code snippets for analysis. For each snippet, we capture the model’s logit outputs at the insertion point and retain the top 30 ranked tokens for further examination.

\subsubsection{Results}

\begin{figure*}[!t]
	\centering
\includegraphics[width=0.9\textwidth]{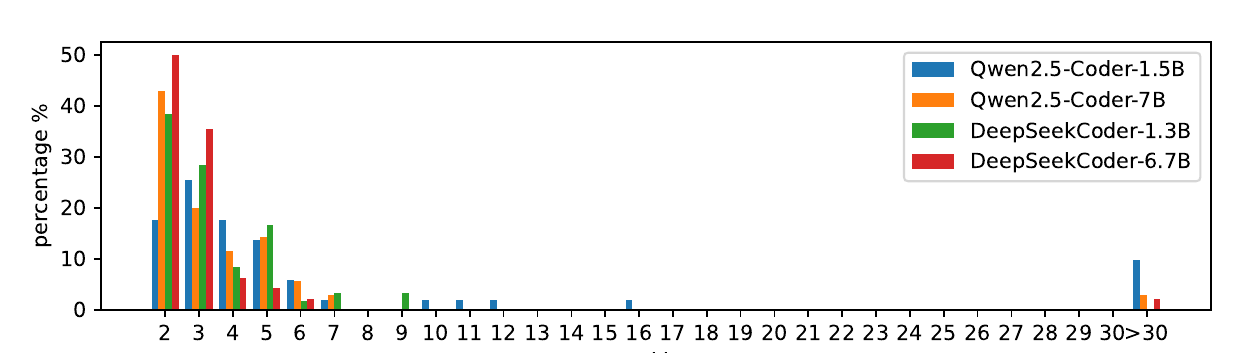}
	\caption{``If'' token ranking}
        \label{ranking}
\end{figure*}
Figure~\ref{ranking} shows that at positions where an ``if'' statement is expected, the ``if'' token consistently ranks within the top five. For \qwencoderbig, when the ``if'' statement is missing, 42.9\% of the cases rank ``if'' second, 20.0\% rank it third, and 11.4\% rank it fourth. Similarly, for \deepseekbig, 50.0\% rank ``if'' second, 35.4\% third, and 6.3\% fourth.

These results suggest that although the model often assigns a high probability to the ``if'' token at critical positions, it occasionally fails to generate it under greedy decoding. This gap indicates that the model inherently recognizes the need for condition statements but that its current sampling strategy may not always capture this potential. Adjusting the sampling strategy—for example, by employing temperature sampling, top-k, or top-p methods—could allow the model to better utilize this latent capability, thereby enhancing overall code robustness.

\finding{The high ranking of the ``if'' token (usually second or third) at positions lacking conditionals suggests that LLMs inherently recognize essential control structures. Optimizing the sampling strategy could help generate more robust code by adding missing ``if'' branches.}

\section{Plug-and-Play Framework for Enhancing Code Robustness}
\label{sec:app}
\begin{figure*}[!t]
	\centering
\includegraphics[width=0.9\textwidth]{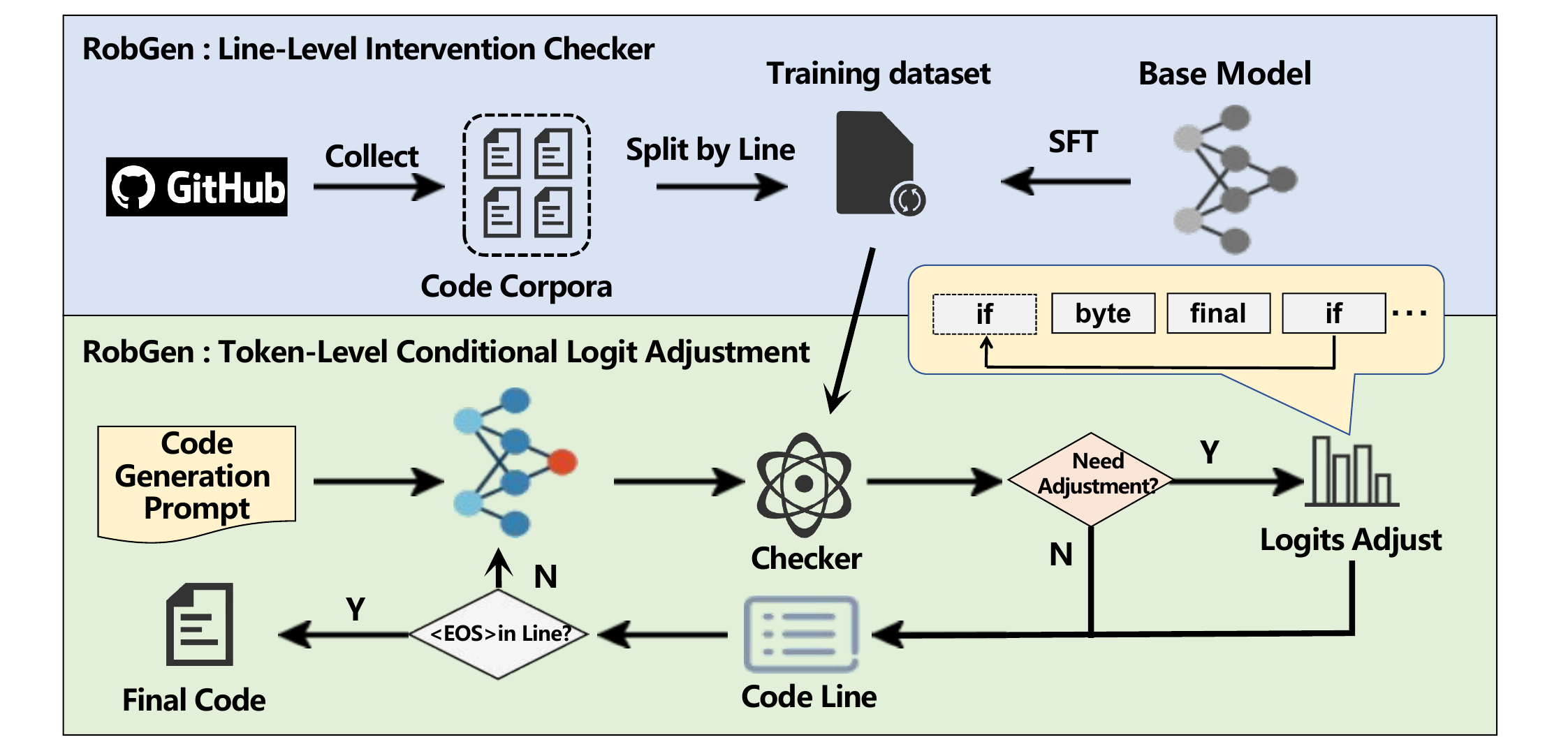}
  	\caption{Overview of \app Framework}
        \label{framework}
\end{figure*}

Building on our empirical findings, we present \app, a plug-and-play framework that can be seamlessly integrated into the decoding process of LLM-based code generation to improve the robustness of the generated code. The underlying model still performs standard next-token prediction in an auto-regressive manner. \app intervenes at the beginning of each line—specifically, at the first non-empty token—where a checker examines the previously generated context and decides whether logit adjustment is required. If adjustment is necessary, the strategy is applied before decoding continues. This procedure repeats until an <EOS> token is generated or the maximum token limit is reached. Figure~\ref{framework} illustrates the overall architecture of \app, which consists of two key components: (i) a checker that determines when intervention is needed (Section~\ref{sec:apptime}), and (ii) a logit adjustment module that specifies how the intervention is applied (Section~\ref{sec:appadjust}). Together, these components address common robustness issues, particularly missing conditional checks, which account for over 90\% of the observed failures (Finding 2).

\subsection{Line-Level Intervention Checker}
\label{sec:apptime}
Uncontrolled logit adjustment at every decoding step may introduce redundant or nested conditional branches. 
To address this, we design a \textbf{line-level intervention checker} that determines whether the upcoming line should begin with an ``if'' statement. 
We adopt a line-level granularity because intervening at every token is both unnecessary and can disrupt normal code generation. 
By focusing on the first non-whitespace token of each line, the checker guides the model to insert essential control structures, 
thereby improving the robustness of the generated code without excessive interference. 
Before generating each new line, the checker receives the previously generated code, including the method signature, and predicts whether a conditional statement is required. 
If the prediction is positive, the logit adjustment strategy described in Section~\ref{sec:appadjust} is applied to the first non-whitespace token; otherwise, decoding proceeds without intervention.

To train the checker effectively, we constructed a dataset to teach it when a new line should initiate a conditional statement. 

\parabf{Training Data.}
We selected three Java repositories from GitHub with over 3,000 stars \cite{li2024devevalmanuallyannotatedcodegeneration}to ensure that the collected methods are of high quality and representative of real-world coding practices. 
All methods were extracted and segmented into code blocks paired with their method signatures. 
To provide supervision for the checker, we labeled code blocks as follows: a block is considered a \textbf{positive sample} if the next line in the original code begins with an ``if'' statement, and a \textbf{negative sample} if it does not. In this context, every code block is required to start with a method signature.
This labeling strategy allows the checker to learn when a conditional statement is required at the beginning of a new line. 
Given that positive examples are naturally less frequent and we aim to avoid unnecessary interventions during decoding, we intentionally include more negative samples. 
To reduce bias and improve training stability, we applied an imbalanced sampling strategy, resulting in 12,000 samples (4,000 positive, 8,000 negative).

\parabf{Training Setup.}
For efficient and effective training, we used Qwen3.0-0.6B \cite{yang2025qwen3technicalreport} as the base model and fine-tuned it with LoRA \cite{hu2021loralowrankadaptationlarge} using supervised fine-tuning (SFT) at a learning rate of 2e-5 for five epochs. The model was evaluated on a validation set to monitor generalization performance. The checkpoint with the highest validation accuracy was retained as the final LoRA weights. 
This setup balances computational efficiency with model adaptation capacity, ensuring the checker can accurately predict when conditional statements should be inserted while keeping training overhead manageable. 
Other hyperparameters followed the configuration in \cite{Robgen}.

\subsection{Token-Level Conditional Logit Adjustment}
\label{sec:appadjust}
Motivated by empirical findings from RQ4 (Section~\ref{sec:rq4}), we observe that LLMs generally rank the ``if'' token highly at positions where a conditional statement is required. 
Since token selection depends on logits, we propose dynamically adjusting the logits of the ``if'' token to guide the model toward generating missing conditional statements, thereby enhancing code robustness. 

We implement a selective logit adjustment mechanism for the ``if'' token. By boosting the logit of ``if'' at strategic positions, the model is guided to generate necessary conditional statements without interfering with overall code synthesis. 
The mechanism is applied during decoding using the \texttt{LogitsProcessor} interface from Transformers \cite{Wolf_Transformers_State-of-the-Art_Natural_2020}.

\parabf{Selective Intervention Rule.}
Unconditional logit adjustment at every decoding step may disrupt normal code generation. 
Based on insights from RQ4 (Finding 4), the ``if'' token typically ranks second or third at positions where a conditional statement is missing. 
Therefore, we apply logit adjustment only when the ``if'' token is among the top three ranked tokens, ensuring targeted and effective intervention.

\parabf{If Token Logit Adjustment.}
To ensure that the ``if'' token’s logit value surpasses competing tokens, we employ the equation in Eq.~\ref{eq:adjst} to adjust the logit value for the ``if'' token.
\begin{equation}
    \label{eq:adjst}
    \text{logit} = \text{logit} + \Delta \times (\text{rank} - 1) 
\end{equation}
where \( \Delta \) denotes the adjustment factor, and \( \text{rank} \) represents the position of the ``if'' token in the logits ranking.

For example, an ``if'' token can be ranked first by suppressing a competing token (e.g., ``byte'') that initially ranks higher with a higher logit value. Note that if the LLM predicts that the ``if'' token has a low probability of being generated based on the context, the adjustment will not affect the result.

To determine \( \Delta \), we analyze RQ4 data by calculating the relative difference between the ``if'' token’s logit and the highest-ranked token, normalized by their rank difference. The adjustment factor \( \Delta \) is set to the 90th-percentile of these observed differences, ensuring it surpasses 90\% of the values. The derived \( \Delta \) values for different models are 1.60, 1.70, 1.29, and 1.78 for \deepseeksmall, \deepseekbig, \qwencodersmall, and \qwencoderbig, respectively. For models not investigated, \( \Delta \) can be set to 1.0.

\section{EVALUATION}
In this section, we focus on the following RQs:

\begin{itemize}[left=1mm, labelsep=1em,itemsep=2pt,topsep=0pt,parsep=0pt]
    \item \textbf{RQ5:} How do different methods perform across various models, and what are the key differences among them?
    \item \textbf{RQ6:} How does the efficiency of different generation methods compare to the default code generation process of the model?   
\end{itemize}

\subsection{RQ5: Method Comparison}
\label{sec:rq5}
\subsubsection{Design}
We conducted experiments using our framework on the four models from our empirical analysis, as well as on \starcoderbig. 
To assess the effectiveness of our approach, we consider the following baseline methods.
\begin{itemize}[left=1mm, labelsep=1em,itemsep=2pt,topsep=0pt,parsep=0pt]

\item \textbf{Greedy Sampling: } The model’s default code generation approach.

\item \textbf{Robust Coder Prompt: } Robust Coder Prompt (RP) involves modifying the prompt to encourage the model to generate more robust code by explicitly guiding it toward incorporating necessary checks and handling potential exceptions. These robustness requirements are embedded into the model's code-generation template, as detailed in \cite{Robgen}.

\item \textbf {Post Generation Insertion:} Based on RQ3 (Section~\ref{sec:rq3}), We propose a post-generation insertion (PGI) mechanism to address robustness issues in the first line of generated code. Missing ``if'' statements are reconstructed using a Fill-in-the-Middle (FIM) prompt~\cite{FIM2022}, guiding the model to generate the necessary conditional logic.
\item \textbf{RobGen without Checker :}When the Checker is not used, the model performs token-level conditional logit adjustment at every token generation step.

\end{itemize}

In all methods, we default to employing greedy sampling, with the token limit set to 1024. All metrics are computed over the full dataset; this differs from the configuration used in previous empirical research.

For different models, the adjustment factor \( \Delta \) used during generation varied. Specifically, for the four models from our empirical analysis, we adopted empirically determined values, whereas for the \starcoderbig , we set the step size to 1.

\subsubsection{ Results}
The metric calculation results, presented in Table \ref{tab:RQ5}, demonstrate that \app effectively enhances the robustness of generated code across five models. While \app achieves the highest average Pass@1(43.57) and average AvgABE(2.42) slightly lower than \app without Checker, it demonstrates superior stability, maintaining the model’s Pass@1 with minimal fluctuation.
When applying \app, the AvgABE of model-generated code exceeds that of the reference code. 
For instance, using \app on \deepseekbig increases the AvgABE from 1.47 to 2.15, while applying \app to \qwencodersmall raises the AvgABE from 1.38 to 2.42—both surpassing the reference code’s AvgABE of 2.03. Notably, \starcoderbig achieves a very high AvgABE score (7.76) under the \app approach without the checker, indicating that \starcoderbig generates numerous redundant ``if'' conditional expressions, which in turn negatively impact the functional correctness of the code. In contrast, when the checker is employed in the \app approach, this abnormal behavior of \starcoderbig is mitigated—or even eliminated—demonstrating that the checker imposes appropriate constraints on the logit adjustments.

\begin{figure*}[!t]
	\centering
\includegraphics[width=0.9\textwidth]{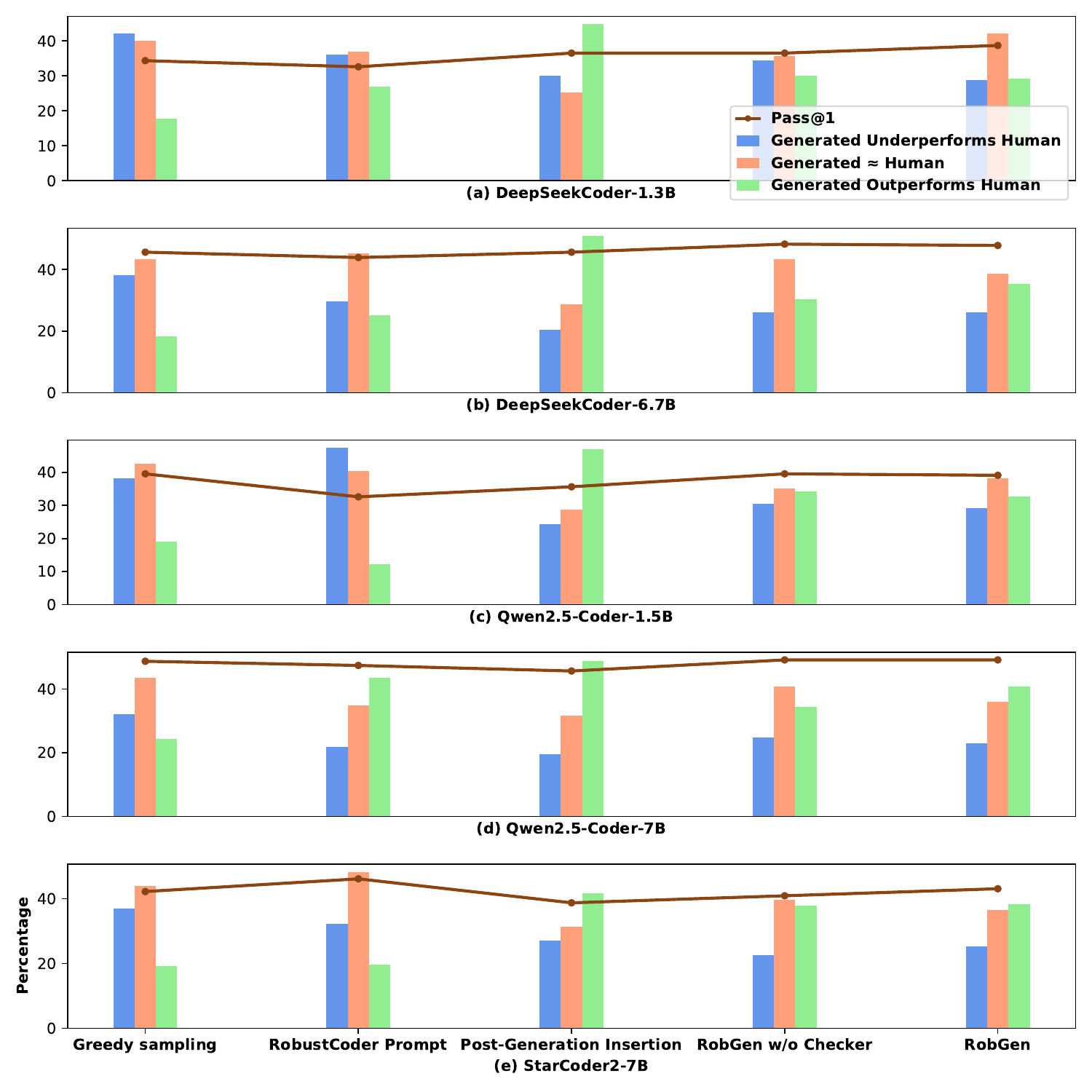}
	\caption{Robustness Evaluation Results and Pass@1 of Different Methods.}
        \label{RQ5_RRI}
\end{figure*}

The Robustness Evaluation Results and Pass@1 distributions in Figure \ref{RQ5_RRI} highlight the effectiveness of \app. For instance, with \app, the proportion of generated code underperform Human for \deepseeksmall drops from 42.2\% to 28.7\%. Similarly, for \starcoderbig, the proportion of generated code outperform Human code rises from 19.1\% to 38.3\% using \app.

In comparison, RobustCoder Prompt (RP) is less effective than \app. Although RP can improve code robustness, its effects are inconsistent and sometimes detrimental—for example, \qwencodersmall sees AvgABE drop from 1.38 to 1.05 and the proportion of code outperforming human-written code fall from 19.1\% to 12.2\%, indicating that prompt adjustment alone is insufficient to enhance robustness.
While Post-Generation Insertion (PGI) improves code robustness, its impact on Pass@1 is inconsistent (e.g., \starcoderbig: 42.2 → 38.7; \qwencodersmall: 48.7 → 45.7). In contrast, \app enhances robustness without compromising Pass@1 and can even improve it (e.g., \deepseeksmall: 34.4 → 38.7; \deepseekbig: 45.7 → 47.8). Without the checker, \app may reduce Pass@1 (e.g., \starcoderbig: 42.2 → 40.9), underscoring the checker's effectiveness.

\begin{table*}[]
\centering
\caption{AvgABE and Pass@1 of Diferent Method: ``GS'' indicates Greedy Sampling, ``RP'' indicates RobustCoder Prompt and ``PGI'' indicates Post-Generation Insertion }
\label{tab:RQ5}

\resizebox{0.9\textwidth}{!}{ 

\begin{tabular}{|cccccc|}
\hline
\multicolumn{1}{|c|}{\textit{\textbf{Model}}} & \multicolumn{1}{c|}{\textit{\textbf{GS}}} & \multicolumn{1}{c|}{\textit{\textbf{RP}}} & \multicolumn{1}{c|}{\textit{\textbf{PGI}}} & \multicolumn{1}{c|}{\textit{\textbf{RobGen w/o Checker}}} & \textit{\textbf{RobGen}} \\ \hline
\multicolumn{6}{|c|}{AvgABE (Ground Truth AvgABE: 2.03)}                                                                                                                                                                                                                  \\ \hline
\multicolumn{1}{|c|}{DSC-1.3B}                & \multicolumn{1}{c|}{1.47}                 & \multicolumn{1}{c|}{1.83}                 & \multicolumn{1}{c|}{2.43}                  & \multicolumn{1}{c|}{\textbf{2.5}}                         & 2.19                     \\ \hline
\multicolumn{1}{|c|}{DSC-6.7B}                & \multicolumn{1}{c|}{1.47}                 & \multicolumn{1}{c|}{1.85}                 & \multicolumn{1}{c|}{\textbf{2.43}}         & \multicolumn{1}{c|}{2.15}                                 & 2.33                     \\ \hline
\multicolumn{1}{|c|}{QWC-1.5B}                & \multicolumn{1}{c|}{1.38}                 & \multicolumn{1}{c|}{1.05}                 & \multicolumn{1}{c|}{2.26}                  & \multicolumn{1}{c|}{\textbf{2.42}}                        & 2.19                     \\ \hline
\multicolumn{1}{|c|}{QWC-7B}                  & \multicolumn{1}{c|}{1.79}                 & \multicolumn{1}{c|}{2.08}                 & \multicolumn{1}{c|}{\textbf{2.53}}         & \multicolumn{1}{c|}{2.24}                                 & 2.37                     \\ \hline
\multicolumn{1}{|c|}{STC-7B}                  & \multicolumn{1}{c|}{2.13}                 & \multicolumn{1}{c|}{1.85}                 & \multicolumn{1}{c|}{3.07}                  & \multicolumn{1}{c|}{\textbf{7.76}}                        & 3.03                     \\ \hline
\multicolumn{1}{|c|}{Avg}                     & \multicolumn{1}{c|}{1.65}                 & \multicolumn{1}{c|}{1.73}                 & \multicolumn{1}{c|}{2.06}                  & \multicolumn{1}{c|}{\textbf{3.41}}                        & 2.42                     \\ \hline
\multicolumn{6}{|c|}{Pass@1}                                                                                                                                                                                                                                              \\ \hline
\multicolumn{1}{|c|}{DSC-1.3B}                & \multicolumn{1}{c|}{34.35}                & \multicolumn{1}{c|}{33.48}                & \multicolumn{1}{c|}{36.52}                 & \multicolumn{1}{c|}{36.52}                                & \textbf{38.70}           \\ \hline
\multicolumn{1}{|c|}{DSC-6.7B}                & \multicolumn{1}{c|}{45.65}                & \multicolumn{1}{c|}{43.91}                & \multicolumn{1}{c|}{45.65}                 & \multicolumn{1}{c|}{\textbf{48.26}}                       & 47.83                    \\ \hline
\multicolumn{1}{|c|}{QWC-1.5B}                & \multicolumn{1}{c|}{\textbf{39.57}}       & \multicolumn{1}{c|}{32.61}                & \multicolumn{1}{c|}{35.65}                 & \multicolumn{1}{c|}{\textbf{39.57}}                       & 39.13                    \\ \hline
\multicolumn{1}{|c|}{QWC-7B}                  & \multicolumn{1}{c|}{48.70}                & \multicolumn{1}{c|}{47.39}                & \multicolumn{1}{c|}{45.65}                 & \multicolumn{1}{c|}{49.13}                                & \textbf{49.13}           \\ \hline
\multicolumn{1}{|c|}{STC-7B}                  & \multicolumn{1}{c|}{42.17}                & \multicolumn{1}{c|}{\textbf{46.09}}       & \multicolumn{1}{c|}{38.70}                 & \multicolumn{1}{c|}{40.87}                                & 43.04                    \\ \hline
\multicolumn{1}{|c|}{Avg}                     & \multicolumn{1}{c|}{42.09}                & \multicolumn{1}{c|}{40.70}                & \multicolumn{1}{c|}{40.43}                 & \multicolumn{1}{c|}{42.87}                                & \textbf{43.57}           \\ \hline
\end{tabular}
}

\end{table*}

\begin{figure*}[!t]
	\centering
\includegraphics[width=0.9\textwidth]{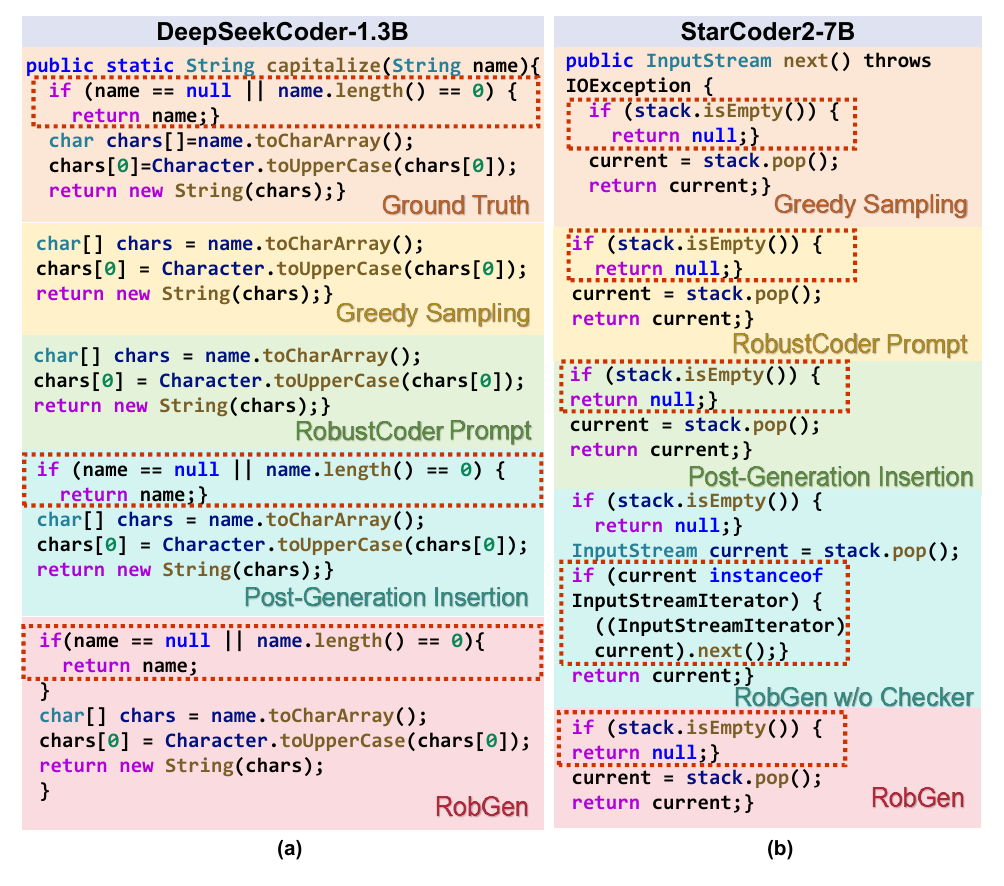}
	\caption{Examples of Code Generated by Different Methods}
        \label{example}
\end{figure*}

Figure \ref{example} presents the generated results of \deepseeksmall  and \starcoderbig using different methods.
In Figure (a), we observe that, compared to the ground truth, the output of Greedy Sampling lacks null pointer and empty array checks. When using RobustCoder Prompt, the model still fails to generate these missing robustness checks. However, when applying PGI and \app, the model successfully generates the necessary robustness checks, as highlighted in the red box. This demonstrates that PGI and \app are more effective than RP in enhancing code robustness.

In Figure (b), We note that when the code is generated by Greedy Sampling, RP, and PGI, it corresponds to the correct version. However, with \app w/o Checker, an unnecessary check of ``\texttt{ current instanceof InputStreamIterator}'' is introduced, leading to redundant and incorrect operational logic. In contrast, when using \app with Checker, the Checker successfully prevents the introduction of such incorrect logic, demonstrating its effective constraining role in logit adjustment.

\finding{\app attains the highest average Pass@1 and offers improved stability, boosting code generation robustness while maintaining or even enhancing the model’s Pass@1, with the Checker serving as a key enforcing mechanism.}

\subsection{RQ6: Efficiency}

\begin{table*}

\caption{Runtime of the model on 20 tasks (Min): ``DSC'' indicates DeepSeekCoder, ``QWC'' indicates Qwen2.5-Coder and ``STC'' indicates StarCoder2-7B.}
\label{runtime}
\resizebox{0.9\textwidth}{!}{ 

\begin{tabular}{|c|c|c|c|c|c|}
\hline
Method             & DSC-1.3B      & DSC-6.7B       & QWC-1.5B                & QWC-7B        & STC-7B         \\ \hline
Greedy Sampling    & 4.67          & 14.09          & 2.41                    & 1.41          & \textbf{14.79} \\ \hline
RobGen w/o Checker & 4.78(+2.3\%)  & 14.16(+0.4\%)  & \textbf{2.74(+13.7\%)}  & 1.45(+1.3\%)  & 14.83(+0.2\%)  \\ \hline
RobGen             & 6.23(+33.4\%) & 15.77(+11.4\%) & \textbf{7.18(+197.9\%)} & 1.65(+17.0\%) & 16.50(+11.6\%) \\ \hline
\end{tabular}%
}
\end{table*}
\subsubsection{Design}
We adopted the experimental setup from RQ5~(Section \ref{sec:rq5}) and randomly selected 20 tasks from the dataset. Each of the five models executed these tasks using our methods, and we recorded the execution time for each model under each method. To mitigate potential errors introduced by external factors, each method for each model was executed three times, and the final execution time was computed as the average of these runs.

\subsubsection{Results}
The execution time results are presented in Table\ref{runtime}. Among all methods, \app exhibits the highest time overhead. For instance, after applying \app, \deepseeksmall’s execution time increased from 4.67 to 6.23, representing a 33.4\% increase compared to Greedy Sampling. 


In contrast, \app w/o Checker has minimal impact on execution time. For example, when using \app w/o Checker, \deepseekbig experienced only a 0.4\% increase, while \starcoderbig saw a negligible 0.2\% increase. 
Although this trade-off is acceptable given that \app requires an additional model invocation, we note that applying \app increases the time overhead of \qwencodersmall to nearly twice that of Greedy Sampling. In contrast, \app w/o Checker introduces minimal overhead. This is because the checker operates at the line level, making it incompatible with KV cache acceleration\cite{yang2025kvlinkacceleratinglargelanguage} and thus significantly increasing inference time for longer code. Excessive code generation by \qwencodersmall leads to longer preceding code, which in turn increases the Checker’s inference time. 

\finding{\app adds only a modest time overhead (+33.4\%), highlighting its efficiency, whereas the Checker’s overhead can rise sharply (+197.9\%) due to model over-generation (repetition), occasionally resulting in abnormal behavior.}


\section{THREATS TO VALIDITY}
Internally, a potential threat is the use of LLMs in our research. To mitigate biases, we use officially released models, deploy them per publishers' guidelines, and validate outputs through tests. We also ensure fairness by using consistent prompts and parameters. Another concern is the representativeness and quality of code data. To address this, we use the CoderEval benchmark for realistic code generation. Due to resource constraints, we evaluated only four models and focused on Java, limiting generalizability. More extensive studies are needed. While RobGen was designed based on findings from these four models, we included Starcoder in experiments to demonstrate broader applicability. 
To minimize human bias in robustness evaluation, we use an LLM Evaluator instead of manual judgment. 
Each task is evaluated three times with randomized input order of human-written and model-generated code to reduce LLM-induced randomness. We tested two prompt designs—pairwise comparison and single-score rating—and found pairwise comparison best aligned with human judgments, which we adopted as our primary evaluation method.



\section{RELATED WORK}
\subsection{LLM-based Code Generation}
LLMs have shown remarkable performance in code generation, with applications in automated code generation \cite{yuan2023evaluating,ugare2024improving,su2024distilled,liu2024stallboostingllmbased,wang2024teachingcodellmsuse}, code translation \cite{yan2023codetransocean,Yang2024Exploring,ou2025repositorylevelcodetranslationbenchmark,wang2024repotransbenchrealworldbenchmarkrepositorylevel}, commit message generation \cite{TaoWei2021Eval,Yichen2023Modification}, unit test generation \cite{MaxandNadi2024Unit,lops2024automatedunittestgeneration,YuanEvaluatingandImproving2024}, and defect localization \cite{an-etal-2024-code,ji2024impact,du2024vulragenhancingllm}. Code-specific LLMs, like DeepSeekCoder \cite{guo2024deepseekcoder}, StarCoder2 \cite{starcoder22024}, and Qwen2.5Coder \cite{qwen2.5}, excel in code-related tasks, with high-quality benchmarks such as HumanEval \cite{chen2021evaluating}, MBPP \cite{austin2021program}, Classeval \cite{DuClassLevelCode2024} and CodeEval \cite{yu2023codereval} assessing their capabilities. While most studies focus on improving code generation and evaluating quality with the Pass@k metric \cite{chen2021evaluating}, some explore code security \cite{li2024llm,islam2024enhancing}, robustness\cite{sarker2024syntacticrobustnessllmbasedcode,awal2024comparingrobustnessadversarialattacks}, hallucination\cite{zhang2025llmhallucinationspracticalcode} and trustworthy\cite{wang2024trustworthyllmscodedatacentric}. However, existing benchmarks fall short in assessing the robustness of model-generated code. This work compares the robustness of LLM-generated code with human-written code, identifying common deficiencies and offering insights for future improvements.

\subsection{Robustness of LLM-generated Code}

Earlier research on LLMs focused primarily on accuracy in small-scale tasks, often overlooking code robustness in real-world development. Recently, studies like Liu et al.\cite{LiuYue2024Quality} have analyzed ChatGPT-generated code for correctness and quality issues, while Zhong et al.\cite{Zhong_Wang_2024} introduced RobustAPI to assess the robustness of LLM-generated code for Java API misuse from Stack Overflow questions. Evaluation results on GPT-4  revealed that 62\% of the generated code still exhibited API misuse issues~\cite{Zhong_Wang_2024}. Recently, Zhang et al. \cite{zhang2024seekerenhancingexceptionhandling} introduced Seeker, a multi-agent framework for improving exception-handling and robustness through automated detection and optimization with a high cost though.

Previous work lacks in-depth analysis of robustness issues in LLM-generated code, particularly in real-world tasks. Many methods are either impractical or involve high overhead. In contrast, our study provides a deeper empirical analysis and identifies common robustness issues across multiple models. We propose \app, a model-agnostic plug-in framework that improves robustness with minimal overhead, enhancing code quality during decoding in a lightweight, adaptable manner.

\section{CONCLUSION}
In conclusion, this study emphasizes the need for robustness in LLM-generated code, which is often overlooked in favor of correctness. Using the CoderEval benchmark, we identify critical gaps, such as missing null checks. To address these, we propose RobGen, a plug-in framework that enhances code robustness without retraining, through techniques like Decoding Adjustment. Our experiments show a 10\% reduction in less robust code, demonstrating the effectiveness of this approach. RobGen offers a flexible, model-agnostic solution for improving LLM-generated code reliability across various tasks.

\begin{acks}
\end{acks}

\bibliographystyle{ACM-Reference-Format}
\bibliography{ref}


\begin{thebibliography}{69}


\ifx \showCODEN    \undefined \def \showCODEN     #1{\unskip}     \fi
\ifx \showDOI      \undefined \def \showDOI       #1{#1}\fi
\ifx \showISBNx    \undefined \def \showISBNx     #1{\unskip}     \fi
\ifx \showISBNxiii \undefined \def \showISBNxiii  #1{\unskip}     \fi
\ifx \showISSN     \undefined \def \showISSN      #1{\unskip}     \fi
\ifx \showLCCN     \undefined \def \showLCCN      #1{\unskip}     \fi
\ifx \shownote     \undefined \def \shownote      #1{#1}          \fi
\ifx \showarticletitle \undefined \def \showarticletitle #1{#1}   \fi
\ifx \showURL      \undefined \def \showURL       {\relax}        \fi
\providecommand\bibfield[2]{#2}
\providecommand\bibinfo[2]{#2}
\providecommand\natexlab[1]{#1}
\providecommand\showeprint[2][]{arXiv:#2}

\bibitem[Rob(2025)]%
        {Robgen}
 \bibinfo{year}{2025}\natexlab{}.
\newblock
\urldef\tempurl%
\url{https://anonymous.4open.science/r/RobGen-838C}
\showURL{%
\tempurl}


\bibitem[An et~al\mbox{.}(2024)]%
        {an-etal-2024-code}
\bibfield{author}{\bibinfo{person}{Jimin An}, \bibinfo{person}{YunSeok Choi}, {and} \bibinfo{person}{Jee-Hyong Lee}.} \bibinfo{year}{2024}\natexlab{}.
\newblock \showarticletitle{Code Defect Detection Using Pre-trained Language Models with Encoder-Decoder via Line-Level Defect Localization}. In \bibinfo{booktitle}{\emph{Proceedings of the 2024 Joint International Conference on Computational Linguistics, Language Resources and Evaluation (LREC-COLING 2024)}}, \bibfield{editor}{\bibinfo{person}{Nicoletta Calzolari}, \bibinfo{person}{Min-Yen Kan}, \bibinfo{person}{Veronique Hoste}, \bibinfo{person}{Alessandro Lenci}, \bibinfo{person}{Sakriani Sakti}, {and} \bibinfo{person}{Nianwen Xue}} (Eds.). \bibinfo{publisher}{ELRA and ICCL}, \bibinfo{address}{Torino, Italia}, \bibinfo{pages}{3446--3456}.
\newblock
\urldef\tempurl%
\url{https://aclanthology.org/2024.lrec-main.306/}
\showURL{%
\tempurl}


\bibitem[Austin et~al\mbox{.}(2021)]%
        {austin2021program}
\bibfield{author}{\bibinfo{person}{Jacob Austin}, \bibinfo{person}{Augustus Odena}, \bibinfo{person}{Maxwell Nye}, \bibinfo{person}{Maarten Bosma}, \bibinfo{person}{Henryk Michalewski}, \bibinfo{person}{David Dohan}, \bibinfo{person}{Ellen Jiang}, \bibinfo{person}{Carrie Cai}, \bibinfo{person}{Michael Terry}, \bibinfo{person}{Quoc Le}, {et~al\mbox{.}}} \bibinfo{year}{2021}\natexlab{}.
\newblock \showarticletitle{Program synthesis with large language models}.
\newblock \bibinfo{journal}{\emph{arXiv preprint arXiv:2108.07732}} (\bibinfo{year}{2021}).
\newblock


\bibitem[Awal et~al\mbox{.}(2024)]%
        {awal2024comparingrobustnessadversarialattacks}
\bibfield{author}{\bibinfo{person}{Md~Abdul Awal}, \bibinfo{person}{Mrigank Rochan}, {and} \bibinfo{person}{Chanchal~K. Roy}.} \bibinfo{year}{2024}\natexlab{}.
\newblock \bibinfo{title}{Comparing Robustness Against Adversarial Attacks in Code Generation: LLM-Generated vs. Human-Written}.
\newblock
\newblock
\showeprint[arxiv]{2411.10565}~[cs.SE]
\urldef\tempurl%
\url{https://arxiv.org/abs/2411.10565}
\showURL{%
\tempurl}


\bibitem[Bavarian et~al\mbox{.}(2022)]%
        {FIM2022}
\bibfield{author}{\bibinfo{person}{Mohammad Bavarian}, \bibinfo{person}{Heewoo Jun}, \bibinfo{person}{Nikolas Tezak}, \bibinfo{person}{John Schulman}, \bibinfo{person}{Christine McLeavey}, \bibinfo{person}{Jerry Tworek}, {and} \bibinfo{person}{Mark Chen}.} \bibinfo{year}{2022}\natexlab{}.
\newblock \bibinfo{title}{Efficient Training of Language Models to Fill in the Middle}.
\newblock
\newblock
\showeprint[arxiv]{2207.14255}~[cs.CL]
\urldef\tempurl%
\url{https://arxiv.org/abs/2207.14255}
\showURL{%
\tempurl}


\bibitem[Busker et~al\mbox{.}(2025)]%
        {BUSKER2025101988}
\bibfield{author}{\bibinfo{person}{Tony Busker}, \bibinfo{person}{Sunil Choenni}, {and} \bibinfo{person}{Mortaza~S. Bargh}.} \bibinfo{year}{2025}\natexlab{}.
\newblock \showarticletitle{Exploiting GPT for synthetic data generation: An empirical study}.
\newblock \bibinfo{journal}{\emph{Government Information Quarterly}} \bibinfo{volume}{42}, \bibinfo{number}{1} (\bibinfo{year}{2025}), \bibinfo{pages}{101988}.
\newblock
\showISSN{0740-624X}
\urldef\tempurl%
\url{https://doi.org/10.1016/j.giq.2024.101988}
\showDOI{\tempurl}


\bibitem[Chen et~al\mbox{.}(2021a)]%
        {chen2021codex}
\bibfield{author}{\bibinfo{person}{Mark Chen}, \bibinfo{person}{Jerry Tworek}, \bibinfo{person}{Heewoo Jun}, \bibinfo{person}{Qiming Yuan}, \bibinfo{person}{Henrique~Ponde de Oliveira~Pinto}, \bibinfo{person}{Jared Kaplan}, \bibinfo{person}{Harri Edwards}, \bibinfo{person}{Yuri Burda}, \bibinfo{person}{Nicholas Joseph}, \bibinfo{person}{Greg Brockman}, \bibinfo{person}{Alex Ray}, \bibinfo{person}{Raul Puri}, \bibinfo{person}{Gretchen Krueger}, \bibinfo{person}{Michael Petrov}, \bibinfo{person}{Heidy Khlaaf}, \bibinfo{person}{Girish Sastry}, \bibinfo{person}{Pamela Mishkin}, \bibinfo{person}{Brooke Chan}, \bibinfo{person}{Scott Gray}, \bibinfo{person}{Nick Ryder}, \bibinfo{person}{Mikhail Pavlov}, \bibinfo{person}{Alethea Power}, \bibinfo{person}{Lukasz Kaiser}, \bibinfo{person}{Mohammad Bavarian}, \bibinfo{person}{Clemens Winter}, \bibinfo{person}{Philippe Tillet}, \bibinfo{person}{Felipe~Petroski Such}, \bibinfo{person}{Dave Cummings}, \bibinfo{person}{Matthias Plappert}, \bibinfo{person}{Fotios Chantzis},
  \bibinfo{person}{Elizabeth Barnes}, \bibinfo{person}{Ariel Herbert-Voss}, \bibinfo{person}{William~Hebgen Guss}, \bibinfo{person}{Alex Nichol}, \bibinfo{person}{Alex Paino}, \bibinfo{person}{Nikolas Tezak}, \bibinfo{person}{Jie Tang}, \bibinfo{person}{Igor Babuschkin}, \bibinfo{person}{Suchir Balaji}, \bibinfo{person}{Shantanu Jain}, \bibinfo{person}{William Saunders}, \bibinfo{person}{Christopher Hesse}, \bibinfo{person}{Andrew~N. Carr}, \bibinfo{person}{Jan Leike}, \bibinfo{person}{Josh Achiam}, \bibinfo{person}{Vedant Misra}, \bibinfo{person}{Evan Morikawa}, \bibinfo{person}{Alec Radford}, \bibinfo{person}{Matthew Knight}, \bibinfo{person}{Miles Brundage}, \bibinfo{person}{Mira Murati}, \bibinfo{person}{Katie Mayer}, \bibinfo{person}{Peter Welinder}, \bibinfo{person}{Bob McGrew}, \bibinfo{person}{Dario Amodei}, \bibinfo{person}{Sam McCandlish}, \bibinfo{person}{Ilya Sutskever}, {and} \bibinfo{person}{Wojciech Zaremba}.} \bibinfo{year}{2021}\natexlab{a}.
\newblock \showarticletitle{Evaluating Large Language Models Trained on Code}.
\newblock  (\bibinfo{year}{2021}).
\newblock
\showeprint[arxiv]{2107.03374}~[cs.LG]


\bibitem[Chen et~al\mbox{.}(2021b)]%
        {chen2021evaluating}
\bibfield{author}{\bibinfo{person}{Mark Chen}, \bibinfo{person}{Jerry Tworek}, \bibinfo{person}{Heewoo Jun}, \bibinfo{person}{Qiming Yuan}, \bibinfo{person}{Henrique Ponde de~Oliveira Pinto}, \bibinfo{person}{Jared Kaplan}, \bibinfo{person}{Harri Edwards}, \bibinfo{person}{Yuri Burda}, \bibinfo{person}{Nicholas Joseph}, \bibinfo{person}{Greg Brockman}, {et~al\mbox{.}}} \bibinfo{year}{2021}\natexlab{b}.
\newblock \showarticletitle{Evaluating large language models trained on code}.
\newblock \bibinfo{journal}{\emph{arXiv preprint arXiv:2107.03374}} (\bibinfo{year}{2021}).
\newblock


\bibitem[Cheng et~al\mbox{.}(2023)]%
        {cheng2023gpt4gooddataanalyst}
\bibfield{author}{\bibinfo{person}{Liying Cheng}, \bibinfo{person}{Xingxuan Li}, {and} \bibinfo{person}{Lidong Bing}.} \bibinfo{year}{2023}\natexlab{}.
\newblock \bibinfo{title}{Is GPT-4 a Good Data Analyst?}
\newblock
\newblock
\showeprint[arxiv]{2305.15038}~[cs.CL]
\urldef\tempurl%
\url{https://arxiv.org/abs/2305.15038}
\showURL{%
\tempurl}


\bibitem[Cohen(1960)]%
        {CohenKappa}
\bibfield{author}{\bibinfo{person}{Jacob Cohen}.} \bibinfo{year}{1960}\natexlab{}.
\newblock \showarticletitle{A Coefficient of Agreement for Nominal Scales}.
\newblock \bibinfo{journal}{\emph{Educational and Psychological Measurement}} \bibinfo{volume}{20}, \bibinfo{number}{1} (\bibinfo{year}{1960}), \bibinfo{pages}{37--46}.
\newblock
\urldef\tempurl%
\url{https://doi.org/10.1177/001316446002000104}
\showDOI{\tempurl}
\showeprint{https://doi.org/10.1177/001316446002000104}


\bibitem[DeepSeek-AI(2024)]%
        {deepseek-llm}
\bibfield{author}{\bibinfo{person}{DeepSeek-AI}.} \bibinfo{year}{2024}\natexlab{}.
\newblock \showarticletitle{DeepSeek LLM: Scaling Open-Source Language Models with Longtermism}.
\newblock \bibinfo{journal}{\emph{arXiv preprint arXiv:2401.02954}} (\bibinfo{year}{2024}).
\newblock
\urldef\tempurl%
\url{https://github.com/deepseek-ai/DeepSeek-LLM}
\showURL{%
\tempurl}


\bibitem[Dong et~al\mbox{.}(2023)]%
        {dong2023self}
\bibfield{author}{\bibinfo{person}{Yihong Dong}, \bibinfo{person}{Xue Jiang}, \bibinfo{person}{Zhi Jin}, {and} \bibinfo{person}{Ge Li}.} \bibinfo{year}{2023}\natexlab{}.
\newblock \showarticletitle{Self-collaboration Code Generation via ChatGPT}.
\newblock \bibinfo{journal}{\emph{arXiv preprint arXiv:2304.07590}} (\bibinfo{year}{2023}).
\newblock


\bibitem[Du et~al\mbox{.}(2024a)]%
        {DuClassLevelCode2024}
\bibfield{author}{\bibinfo{person}{Xueying Du}, \bibinfo{person}{Mingwei Liu}, \bibinfo{person}{Kaixin Wang}, \bibinfo{person}{Hanlin Wang}, \bibinfo{person}{Junwei Liu}, \bibinfo{person}{Yixuan Chen}, \bibinfo{person}{Jiayi Feng}, \bibinfo{person}{Chaofeng Sha}, \bibinfo{person}{Xin Peng}, {and} \bibinfo{person}{Yiling Lou}.} \bibinfo{year}{2024}\natexlab{a}.
\newblock \showarticletitle{Evaluating Large Language Models in Class-Level Code Generation} \emph{(\bibinfo{series}{ICSE '24})}. \bibinfo{publisher}{Association for Computing Machinery}, \bibinfo{address}{New York, NY, USA}, Article \bibinfo{articleno}{81}, \bibinfo{numpages}{13}~pages.
\newblock
\showISBNx{9798400702174}
\urldef\tempurl%
\url{https://doi.org/10.1145/3597503.3639219}
\showDOI{\tempurl}


\bibitem[Du et~al\mbox{.}(2024b)]%
        {du2024vulragenhancingllm}
\bibfield{author}{\bibinfo{person}{Xueying Du}, \bibinfo{person}{Geng Zheng}, \bibinfo{person}{Kaixin Wang}, \bibinfo{person}{Jiayi Feng}, \bibinfo{person}{Wentai Deng}, \bibinfo{person}{Mingwei Liu}, \bibinfo{person}{Bihuan Chen}, \bibinfo{person}{Xin Peng}, \bibinfo{person}{Tao Ma}, {and} \bibinfo{person}{Yiling Lou}.} \bibinfo{year}{2024}\natexlab{b}.
\newblock \bibinfo{title}{Vul-RAG: Enhancing LLM-based Vulnerability Detection via Knowledge-level RAG}.
\newblock
\newblock
\showeprint[arxiv]{2406.11147}~[cs.SE]
\urldef\tempurl%
\url{https://arxiv.org/abs/2406.11147}
\showURL{%
\tempurl}


\bibitem[Goodfellow et~al\mbox{.}(2016)]%
        {Goodfellow2016}
\bibfield{author}{\bibinfo{person}{Ian Goodfellow}, \bibinfo{person}{Yoshua Bengio}, {and} \bibinfo{person}{Aaron Courville}.} \bibinfo{year}{2016}\natexlab{}.
\newblock \bibinfo{booktitle}{\emph{Deep Learning}}.
\newblock \bibinfo{publisher}{MIT Press}.
\newblock


\bibitem[Guo et~al\mbox{.}(2024b)]%
        {guo2024deepseek}
\bibfield{author}{\bibinfo{person}{Daya Guo}, \bibinfo{person}{Qihao Zhu}, \bibinfo{person}{Dejian Yang}, \bibinfo{person}{Zhenda Xie}, \bibinfo{person}{Kai Dong}, \bibinfo{person}{Wentao Zhang}, \bibinfo{person}{Guanting Chen}, \bibinfo{person}{Xiao Bi}, \bibinfo{person}{Y Wu}, \bibinfo{person}{YK Li}, {et~al\mbox{.}}} \bibinfo{year}{2024}\natexlab{b}.
\newblock \showarticletitle{DeepSeek-Coder: When the Large Language Model Meets Programming--The Rise of Code Intelligence}.
\newblock \bibinfo{journal}{\emph{arXiv preprint arXiv:2401.14196}} (\bibinfo{year}{2024}).
\newblock


\bibitem[Guo et~al\mbox{.}(2024c)]%
        {guo2024deepseekcoder}
\bibfield{author}{\bibinfo{person}{Daya Guo}, \bibinfo{person}{Qihao Zhu}, \bibinfo{person}{Dejian Yang}, \bibinfo{person}{Zhenda Xie}, \bibinfo{person}{Kai Dong}, \bibinfo{person}{Wentao Zhang}, \bibinfo{person}{Guanting Chen}, \bibinfo{person}{Xiao Bi}, \bibinfo{person}{Y Wu}, \bibinfo{person}{YK Li}, {et~al\mbox{.}}} \bibinfo{year}{2024}\natexlab{c}.
\newblock \showarticletitle{DeepSeek-Coder: When the Large Language Model Meets Programming--The Rise of Code Intelligence}.
\newblock \bibinfo{journal}{\emph{arXiv preprint arXiv:2401.14196}} (\bibinfo{year}{2024}).
\newblock


\bibitem[Guo et~al\mbox{.}(2024a)]%
        {WhentoStop2024}
\bibfield{author}{\bibinfo{person}{Lianghong Guo}, \bibinfo{person}{Yanlin Wang}, \bibinfo{person}{Ensheng Shi}, \bibinfo{person}{Wanjun Zhong}, \bibinfo{person}{Hongyu Zhang}, \bibinfo{person}{Jiachi Chen}, \bibinfo{person}{Ruikai Zhang}, \bibinfo{person}{Yuchi Ma}, {and} \bibinfo{person}{Zibin Zheng}.} \bibinfo{year}{2024}\natexlab{a}.
\newblock \showarticletitle{When to Stop? Towards Efficient Code Generation in LLMs with Excess Token Prevention}. In \bibinfo{booktitle}{\emph{Proceedings of the 33rd ACM SIGSOFT International Symposium on Software Testing and Analysis}} (Vienna, Austria) \emph{(\bibinfo{series}{ISSTA 2024})}. \bibinfo{publisher}{Association for Computing Machinery}, \bibinfo{address}{New York, NY, USA}, \bibinfo{pages}{1073–1085}.
\newblock
\showISBNx{9798400706127}
\urldef\tempurl%
\url{https://doi.org/10.1145/3650212.3680343}
\showDOI{\tempurl}


\bibitem[He et~al\mbox{.}(2023)]%
        {Yichen2023Modification}
\bibfield{author}{\bibinfo{person}{Yichen He}, \bibinfo{person}{Liran Wang}, \bibinfo{person}{Kaiyi Wang}, \bibinfo{person}{Yupeng Zhang}, \bibinfo{person}{Hang Zhang}, {and} \bibinfo{person}{Zhoujun Li}.} \bibinfo{year}{2023}\natexlab{}.
\newblock \showarticletitle{COME: Commit Message Generation with Modification Embedding}. In \bibinfo{booktitle}{\emph{Proceedings of the 32nd ACM SIGSOFT International Symposium on Software Testing and Analysis}} (Seattle, WA, USA) \emph{(\bibinfo{series}{ISSTA 2023})}. \bibinfo{publisher}{Association for Computing Machinery}, \bibinfo{address}{New York, NY, USA}, \bibinfo{pages}{792–803}.
\newblock
\showISBNx{9798400702211}
\urldef\tempurl%
\url{https://doi.org/10.1145/3597926.3598096}
\showDOI{\tempurl}


\bibitem[Hendrik~van Antwerpen(2022)]%
        {treesiter}
\bibfield{author}{\bibinfo{person}{Douglas~A.Creager Hendrik~van Antwerpen, Rob~Rix}.} \bibinfo{year}{2022}\natexlab{}.
\newblock \bibinfo{title}{tree-sitter-graph}.
\newblock \bibinfo{howpublished}{\url{https://github.com/tree-sitter/tree-sitter}}.
\newblock


\bibitem[Hu et~al\mbox{.}(2021)]%
        {hu2021loralowrankadaptationlarge}
\bibfield{author}{\bibinfo{person}{Edward~J. Hu}, \bibinfo{person}{Yelong Shen}, \bibinfo{person}{Phillip Wallis}, \bibinfo{person}{Zeyuan Allen-Zhu}, \bibinfo{person}{Yuanzhi Li}, \bibinfo{person}{Shean Wang}, \bibinfo{person}{Lu Wang}, {and} \bibinfo{person}{Weizhu Chen}.} \bibinfo{year}{2021}\natexlab{}.
\newblock \bibinfo{title}{LoRA: Low-Rank Adaptation of Large Language Models}.
\newblock
\newblock
\showeprint[arxiv]{2106.09685}~[cs.CL]
\urldef\tempurl%
\url{https://arxiv.org/abs/2106.09685}
\showURL{%
\tempurl}


\bibitem[Hui et~al\mbox{.}(2024)]%
        {hui2024qwen25codertechnicalreport}
\bibfield{author}{\bibinfo{person}{Binyuan Hui}, \bibinfo{person}{Jian Yang}, \bibinfo{person}{Zeyu Cui}, \bibinfo{person}{Jiaxi Yang}, \bibinfo{person}{Dayiheng Liu}, \bibinfo{person}{Lei Zhang}, \bibinfo{person}{Tianyu Liu}, \bibinfo{person}{Jiajun Zhang}, \bibinfo{person}{Bowen Yu}, \bibinfo{person}{Keming Lu}, \bibinfo{person}{Kai Dang}, \bibinfo{person}{Yang Fan}, \bibinfo{person}{Yichang Zhang}, \bibinfo{person}{An Yang}, \bibinfo{person}{Rui Men}, \bibinfo{person}{Fei Huang}, \bibinfo{person}{Bo Zheng}, \bibinfo{person}{Yibo Miao}, \bibinfo{person}{Shanghaoran Quan}, \bibinfo{person}{Yunlong Feng}, \bibinfo{person}{Xingzhang Ren}, \bibinfo{person}{Xuancheng Ren}, \bibinfo{person}{Jingren Zhou}, {and} \bibinfo{person}{Junyang Lin}.} \bibinfo{year}{2024}\natexlab{}.
\newblock \bibinfo{title}{Qwen2.5-Coder Technical Report}.
\newblock
\newblock
\showeprint[arxiv]{2409.12186}~[cs.CL]
\urldef\tempurl%
\url{https://arxiv.org/abs/2409.12186}
\showURL{%
\tempurl}


\bibitem[Islam et~al\mbox{.}(2024a)]%
        {islam2024map}
\bibfield{author}{\bibinfo{person}{Md.~Ashraful Islam}, \bibinfo{person}{Mohammed~Eunus Ali}, {and} \bibinfo{person}{Md~Rizwan Parvez}.} \bibinfo{year}{2024}\natexlab{a}.
\newblock \bibinfo{title}{MapCoder: Multi-Agent Code Generation for Competitive Problem Solving}.
\newblock
\newblock
\showeprint[arxiv]{2405.11403}~[cs.CL]
\urldef\tempurl%
\url{https://arxiv.org/abs/2405.11403}
\showURL{%
\tempurl}


\bibitem[Islam et~al\mbox{.}(2024b)]%
        {islam2024enhancing}
\bibfield{author}{\bibinfo{person}{Nafis~Tanveer Islam}, \bibinfo{person}{Joseph Khoury}, \bibinfo{person}{Andrew Seong}, \bibinfo{person}{Elias Bou-Harb}, {and} \bibinfo{person}{Peyman Najafirad}.} \bibinfo{year}{2024}\natexlab{b}.
\newblock \showarticletitle{Enhancing Source Code Security with LLMs: Demystifying The Challenges and Generating Reliable Repairs}.
\newblock \bibinfo{journal}{\emph{arXiv preprint arXiv:2409.00571}} (\bibinfo{year}{2024}).
\newblock


\bibitem[Ji et~al\mbox{.}(2024)]%
        {ji2024impact}
\bibfield{author}{\bibinfo{person}{Suhwan Ji}, \bibinfo{person}{Sanghwa Lee}, \bibinfo{person}{Changsup Lee}, \bibinfo{person}{Hyeonseung Im}, {and} \bibinfo{person}{Yo-Sub Han}.} \bibinfo{year}{2024}\natexlab{}.
\newblock \showarticletitle{Impact of Large Language Models of Code on Fault Localization}.
\newblock \bibinfo{journal}{\emph{arXiv preprint arXiv:2408.09657}} (\bibinfo{year}{2024}).
\newblock


\bibitem[Jiang et~al\mbox{.}(2023)]%
        {jiang2023selfevolve}
\bibfield{author}{\bibinfo{person}{Shuyang Jiang}, \bibinfo{person}{Yuhao Wang}, {and} \bibinfo{person}{Yu Wang}.} \bibinfo{year}{2023}\natexlab{}.
\newblock \showarticletitle{Selfevolve: A code evolution framework via large language models}.
\newblock \bibinfo{journal}{\emph{arXiv preprint arXiv:2306.02907}} (\bibinfo{year}{2023}).
\newblock


\bibitem[Li et~al\mbox{.}(2024b)]%
        {li2024devevalmanuallyannotatedcodegeneration}
\bibfield{author}{\bibinfo{person}{Jia Li}, \bibinfo{person}{Ge Li}, \bibinfo{person}{Yunfei Zhao}, \bibinfo{person}{Yongmin Li}, \bibinfo{person}{Huanyu Liu}, \bibinfo{person}{Hao Zhu}, \bibinfo{person}{Lecheng Wang}, \bibinfo{person}{Kaibo Liu}, \bibinfo{person}{Zheng Fang}, \bibinfo{person}{Lanshen Wang}, \bibinfo{person}{Jiazheng Ding}, \bibinfo{person}{Xuanming Zhang}, \bibinfo{person}{Yuqi Zhu}, \bibinfo{person}{Yihong Dong}, \bibinfo{person}{Zhi Jin}, \bibinfo{person}{Binhua Li}, \bibinfo{person}{Fei Huang}, {and} \bibinfo{person}{Yongbin Li}.} \bibinfo{year}{2024}\natexlab{b}.
\newblock \bibinfo{title}{DevEval: A Manually-Annotated Code Generation Benchmark Aligned with Real-World Code Repositories}.
\newblock
\newblock
\showeprint[arxiv]{2405.19856}~[cs.CL]
\urldef\tempurl%
\url{https://arxiv.org/abs/2405.19856}
\showURL{%
\tempurl}


\bibitem[Li et~al\mbox{.}(2023)]%
        {li2023starcoder}
\bibfield{author}{\bibinfo{person}{Raymond Li}, \bibinfo{person}{Loubna~Ben Allal}, \bibinfo{person}{Yangtian Zi}, \bibinfo{person}{Niklas Muennighoff}, \bibinfo{person}{Denis Kocetkov}, \bibinfo{person}{Chenghao Mou}, \bibinfo{person}{Marc Marone}, \bibinfo{person}{Christopher Akiki}, \bibinfo{person}{Jia Li}, \bibinfo{person}{Jenny Chim}, {et~al\mbox{.}}} \bibinfo{year}{2023}\natexlab{}.
\newblock \showarticletitle{Starcoder: may the source be with you!}
\newblock \bibinfo{journal}{\emph{arXiv preprint arXiv:2305.06161}} (\bibinfo{year}{2023}).
\newblock


\bibitem[Li et~al\mbox{.}(2024a)]%
        {li2024llm}
\bibfield{author}{\bibinfo{person}{Ziyang Li}, \bibinfo{person}{Saikat Dutta}, {and} \bibinfo{person}{Mayur Naik}.} \bibinfo{year}{2024}\natexlab{a}.
\newblock \showarticletitle{Llm-assisted static analysis for detecting security vulnerabilities}.
\newblock \bibinfo{journal}{\emph{arXiv preprint arXiv:2405.17238}} (\bibinfo{year}{2024}).
\newblock


\bibitem[Liu et~al\mbox{.}(2024a)]%
        {liu2024stallboostingllmbased}
\bibfield{author}{\bibinfo{person}{Junwei Liu}, \bibinfo{person}{Yixuan Chen}, \bibinfo{person}{Mingwei Liu}, \bibinfo{person}{Xin Peng}, {and} \bibinfo{person}{Yiling Lou}.} \bibinfo{year}{2024}\natexlab{a}.
\newblock \bibinfo{title}{STALL+: Boosting LLM-based Repository-level Code Completion with Static Analysis}.
\newblock
\newblock
\showeprint[arxiv]{2406.10018}~[cs.SE]
\urldef\tempurl%
\url{https://arxiv.org/abs/2406.10018}
\showURL{%
\tempurl}


\bibitem[Liu et~al\mbox{.}(2024c)]%
        {liu2024your}
\bibfield{author}{\bibinfo{person}{Jiawei Liu}, \bibinfo{person}{Chunqiu~Steven Xia}, \bibinfo{person}{Yuyao Wang}, {and} \bibinfo{person}{Lingming Zhang}.} \bibinfo{year}{2024}\natexlab{c}.
\newblock \showarticletitle{Is your code generated by chatgpt really correct? rigorous evaluation of large language models for code generation}.
\newblock \bibinfo{journal}{\emph{Advances in Neural Information Processing Systems}}  \bibinfo{volume}{36} (\bibinfo{year}{2024}).
\newblock


\bibitem[Liu et~al\mbox{.}(2024b)]%
        {LiuYue2024Quality}
\bibfield{author}{\bibinfo{person}{Yue Liu}, \bibinfo{person}{Thanh Le-Cong}, \bibinfo{person}{Ratnadira Widyasari}, \bibinfo{person}{Chakkrit Tantithamthavorn}, \bibinfo{person}{Li Li}, \bibinfo{person}{Xuan-Bach~D. Le}, {and} \bibinfo{person}{David Lo}.} \bibinfo{year}{2024}\natexlab{b}.
\newblock \showarticletitle{Refining ChatGPT-Generated Code: Characterizing and Mitigating Code Quality Issues}.
\newblock \bibinfo{journal}{\emph{ACM Trans. Softw. Eng. Methodol.}} \bibinfo{volume}{33}, \bibinfo{number}{5}, Article \bibinfo{articleno}{116} (\bibinfo{date}{June} \bibinfo{year}{2024}), \bibinfo{numpages}{26}~pages.
\newblock
\showISSN{1049-331X}
\urldef\tempurl%
\url{https://doi.org/10.1145/3643674}
\showDOI{\tempurl}


\bibitem[Lops et~al\mbox{.}(2024)]%
        {lops2024automatedunittestgeneration}
\bibfield{author}{\bibinfo{person}{Andrea Lops}, \bibinfo{person}{Fedelucio Narducci}, \bibinfo{person}{Azzurra Ragone}, \bibinfo{person}{Michelantonio Trizio}, {and} \bibinfo{person}{Claudio Bartolini}.} \bibinfo{year}{2024}\natexlab{}.
\newblock \bibinfo{title}{A System for Automated Unit Test Generation Using Large Language Models and Assessment of Generated Test Suites}.
\newblock
\newblock
\showeprint[arxiv]{2408.07846}~[cs.SE]
\urldef\tempurl%
\url{https://arxiv.org/abs/2408.07846}
\showURL{%
\tempurl}


\bibitem[Lozhkov et~al\mbox{.}(2024)]%
        {starcoder22024}
\bibfield{author}{\bibinfo{person}{Anton Lozhkov}, \bibinfo{person}{Raymond Li}, \bibinfo{person}{Loubna~Ben Allal}, \bibinfo{person}{Federico Cassano}, \bibinfo{person}{Joel Lamy-Poirier}, \bibinfo{person}{Nouamane Tazi}, \bibinfo{person}{Ao Tang}, \bibinfo{person}{Dmytro Pykhtar}, \bibinfo{person}{Jiawei Liu}, \bibinfo{person}{Yuxiang Wei}, \bibinfo{person}{Tianyang Liu}, \bibinfo{person}{Max Tian}, \bibinfo{person}{Denis Kocetkov}, \bibinfo{person}{Arthur Zucker}, \bibinfo{person}{Younes Belkada}, \bibinfo{person}{Zijian Wang}, \bibinfo{person}{Qian Liu}, \bibinfo{person}{Dmitry Abulkhanov}, \bibinfo{person}{Indraneil Paul}, \bibinfo{person}{Zhuang Li}, \bibinfo{person}{Wen-Ding Li}, \bibinfo{person}{Megan Risdal}, \bibinfo{person}{Jia Li}, \bibinfo{person}{Jian Zhu}, \bibinfo{person}{Terry~Yue Zhuo}, \bibinfo{person}{Evgenii Zheltonozhskii}, \bibinfo{person}{Nii Osae~Osae Dade}, \bibinfo{person}{Wenhao Yu}, \bibinfo{person}{Lucas Krauß}, \bibinfo{person}{Naman Jain}, \bibinfo{person}{Yixuan Su},
  \bibinfo{person}{Xuanli He}, \bibinfo{person}{Manan Dey}, \bibinfo{person}{Edoardo Abati}, \bibinfo{person}{Yekun Chai}, \bibinfo{person}{Niklas Muennighoff}, \bibinfo{person}{Xiangru Tang}, \bibinfo{person}{Muhtasham Oblokulov}, \bibinfo{person}{Christopher Akiki}, \bibinfo{person}{Marc Marone}, \bibinfo{person}{Chenghao Mou}, \bibinfo{person}{Mayank Mishra}, \bibinfo{person}{Alex Gu}, \bibinfo{person}{Binyuan Hui}, \bibinfo{person}{Tri Dao}, \bibinfo{person}{Armel Zebaze}, \bibinfo{person}{Olivier Dehaene}, \bibinfo{person}{Nicolas Patry}, \bibinfo{person}{Canwen Xu}, \bibinfo{person}{Julian McAuley}, \bibinfo{person}{Han Hu}, \bibinfo{person}{Torsten Scholak}, \bibinfo{person}{Sebastien Paquet}, \bibinfo{person}{Jennifer Robinson}, \bibinfo{person}{Carolyn~Jane Anderson}, \bibinfo{person}{Nicolas Chapados}, \bibinfo{person}{Mostofa Patwary}, \bibinfo{person}{Nima Tajbakhsh}, \bibinfo{person}{Yacine Jernite}, \bibinfo{person}{Carlos~Muñoz Ferrandis}, \bibinfo{person}{Lingming Zhang},
  \bibinfo{person}{Sean Hughes}, \bibinfo{person}{Thomas Wolf}, \bibinfo{person}{Arjun Guha}, \bibinfo{person}{Leandro von Werra}, {and} \bibinfo{person}{Harm de Vries}.} \bibinfo{year}{2024}\natexlab{}.
\newblock \bibinfo{title}{StarCoder 2 and The Stack v2: The Next Generation}.
\newblock
\newblock
\showeprint[arxiv]{2402.19173}~[cs.SE]
\urldef\tempurl%
\url{https://arxiv.org/abs/2402.19173}
\showURL{%
\tempurl}


\bibitem[McConnell(2004)]%
        {denfensiveprograming}
\bibfield{author}{\bibinfo{person}{Steve McConnell}.} \bibinfo{year}{2004}\natexlab{}.
\newblock \bibinfo{booktitle}{\emph{Code Complete, Second Edition}}.
\newblock \bibinfo{publisher}{Microsoft Press}, \bibinfo{address}{USA}.
\newblock
\showISBNx{0735619670}


\bibitem[Nijkamp et~al\mbox{.}(2022)]%
        {nijkamp2023codegen}
\bibfield{author}{\bibinfo{person}{Erik Nijkamp}, \bibinfo{person}{Bo Pang}, \bibinfo{person}{Hiroaki Hayashi}, \bibinfo{person}{Lifu Tu}, \bibinfo{person}{Huan Wang}, \bibinfo{person}{Yingbo Zhou}, \bibinfo{person}{Silvio Savarese}, {and} \bibinfo{person}{Caiming Xiong}.} \bibinfo{year}{2022}\natexlab{}.
\newblock \showarticletitle{Codegen: An open large language model for code with multi-turn program synthesis}.
\newblock \bibinfo{journal}{\emph{arXiv preprint arXiv:2203.13474}} (\bibinfo{year}{2022}).
\newblock


\bibitem[OpenAI(2021)]%
        {openai2021codex}
\bibfield{author}{\bibinfo{person}{OpenAI}.} \bibinfo{year}{2021}\natexlab{}.
\newblock \bibinfo{title}{OpenAI Code}.
\newblock \bibinfo{howpublished}{\url{https://openai.com/blog/openai-code}}.
\newblock


\bibitem[Ou et~al\mbox{.}(2025)]%
        {ou2025repositorylevelcodetranslationbenchmark}
\bibfield{author}{\bibinfo{person}{Guangsheng Ou}, \bibinfo{person}{Mingwei Liu}, \bibinfo{person}{Yuxuan Chen}, \bibinfo{person}{Xin Peng}, {and} \bibinfo{person}{Zibin Zheng}.} \bibinfo{year}{2025}\natexlab{}.
\newblock \bibinfo{title}{Repository-level Code Translation Benchmark Targeting Rust}.
\newblock
\newblock
\showeprint[arxiv]{2411.13990}~[cs.SE]
\urldef\tempurl%
\url{https://arxiv.org/abs/2411.13990}
\showURL{%
\tempurl}


\bibitem[Ouh et~al\mbox{.}(2023)]%
        {ouh2023chatgpt}
\bibfield{author}{\bibinfo{person}{Eng~Lieh Ouh}, \bibinfo{person}{Benjamin Kok~Siew Gan}, \bibinfo{person}{Kyong Jin~Shim}, {and} \bibinfo{person}{Swavek Wlodkowski}.} \bibinfo{year}{2023}\natexlab{}.
\newblock \showarticletitle{ChatGPT, Can You Generate Solutions for my Coding Exercises? An Evaluation on its Effectiveness in an undergraduate Java Programming Course.}. In \bibinfo{booktitle}{\emph{Proceedings of the 2023 Conference on Innovation and Technology in Computer Science Education V. 1}}. \bibinfo{pages}{54--60}.
\newblock


\bibitem[Pearson(1900)]%
        {Pearson1900}
\bibfield{author}{\bibinfo{person}{Karl Pearson}.} \bibinfo{year}{1900}\natexlab{}.
\newblock \showarticletitle{X. On the criterion that a given system of deviations from the probable in the case of a correlated system of variables is such that it can be reasonably supposed to have arisen from random sampling}.
\newblock \bibinfo{journal}{\emph{The London, Edinburgh, and Dublin Philosophical Magazine and Journal of Science}} \bibinfo{volume}{50}, \bibinfo{number}{302} (\bibinfo{date}{July} \bibinfo{year}{1900}), \bibinfo{pages}{157--175}.
\newblock
\urldef\tempurl%
\url{https://doi.org/10.1080/14786440009463897}
\showDOI{\tempurl}


\bibitem[Radford et~al\mbox{.}(2018)]%
        {radford2018improving}
\bibfield{author}{\bibinfo{person}{Alec Radford}, \bibinfo{person}{Karthik Narasimhan}, \bibinfo{person}{Tim Salimans}, \bibinfo{person}{Ilya Sutskever}, {et~al\mbox{.}}} \bibinfo{year}{2018}\natexlab{}.
\newblock \showarticletitle{Improving language understanding by generative pre-training}.
\newblock  (\bibinfo{year}{2018}).
\newblock


\bibitem[Roziere et~al\mbox{.}(2023a)]%
        {roziere2023code}
\bibfield{author}{\bibinfo{person}{Baptiste Roziere}, \bibinfo{person}{Jonas Gehring}, \bibinfo{person}{Fabian Gloeckle}, \bibinfo{person}{Sten Sootla}, \bibinfo{person}{Itai Gat}, \bibinfo{person}{Xiaoqing~Ellen Tan}, \bibinfo{person}{Yossi Adi}, \bibinfo{person}{Jingyu Liu}, \bibinfo{person}{Tal Remez}, \bibinfo{person}{J{\'e}r{\'e}my Rapin}, {et~al\mbox{.}}} \bibinfo{year}{2023}\natexlab{a}.
\newblock \showarticletitle{Code llama: Open foundation models for code}.
\newblock \bibinfo{journal}{\emph{arXiv preprint arXiv:2308.12950}} (\bibinfo{year}{2023}).
\newblock


\bibitem[Roziere et~al\mbox{.}(2023b)]%
        {rozière2024code}
\bibfield{author}{\bibinfo{person}{Baptiste Roziere}, \bibinfo{person}{Jonas Gehring}, \bibinfo{person}{Fabian Gloeckle}, \bibinfo{person}{Sten Sootla}, \bibinfo{person}{Itai Gat}, \bibinfo{person}{Xiaoqing~Ellen Tan}, \bibinfo{person}{Yossi Adi}, \bibinfo{person}{Jingyu Liu}, \bibinfo{person}{Tal Remez}, \bibinfo{person}{J{\'e}r{\'e}my Rapin}, {et~al\mbox{.}}} \bibinfo{year}{2023}\natexlab{b}.
\newblock \showarticletitle{Code llama: Open foundation models for code}.
\newblock \bibinfo{journal}{\emph{arXiv preprint arXiv:2308.12950}} (\bibinfo{year}{2023}).
\newblock


\bibitem[Sarker et~al\mbox{.}(2024)]%
        {sarker2024syntacticrobustnessllmbasedcode}
\bibfield{author}{\bibinfo{person}{Laboni Sarker}, \bibinfo{person}{Mara Downing}, \bibinfo{person}{Achintya Desai}, {and} \bibinfo{person}{Tevfik Bultan}.} \bibinfo{year}{2024}\natexlab{}.
\newblock \bibinfo{title}{Syntactic Robustness for LLM-based Code Generation}.
\newblock
\newblock
\showeprint[arxiv]{2404.01535}~[cs.SE]
\urldef\tempurl%
\url{https://arxiv.org/abs/2404.01535}
\showURL{%
\tempurl}


\bibitem[Schäfer et~al\mbox{.}(2024)]%
        {MaxandNadi2024Unit}
\bibfield{author}{\bibinfo{person}{Max Schäfer}, \bibinfo{person}{Sarah Nadi}, \bibinfo{person}{Aryaz Eghbali}, {and} \bibinfo{person}{Frank Tip}.} \bibinfo{year}{2024}\natexlab{}.
\newblock \showarticletitle{An Empirical Evaluation of Using Large Language Models for Automated Unit Test Generation}.
\newblock \bibinfo{journal}{\emph{IEEE Transactions on Software Engineering}} \bibinfo{volume}{50}, \bibinfo{number}{1} (\bibinfo{year}{2024}), \bibinfo{pages}{85--105}.
\newblock
\urldef\tempurl%
\url{https://doi.org/10.1109/TSE.2023.3334955}
\showDOI{\tempurl}


\bibitem[St.~Amour and Tilevich(2024)]%
        {st2024toward}
\bibfield{author}{\bibinfo{person}{Leo St.~Amour} {and} \bibinfo{person}{Eli Tilevich}.} \bibinfo{year}{2024}\natexlab{}.
\newblock \showarticletitle{Toward Declarative Auditing of Java Software for Graceful Exception Handling}. In \bibinfo{booktitle}{\emph{Proceedings of the 21st ACM SIGPLAN International Conference on Managed Programming Languages and Runtimes}}. \bibinfo{pages}{90--97}.
\newblock


\bibitem[Strauss and Corbin(1990)]%
        {strauss1990basics}
\bibfield{author}{\bibinfo{person}{Anselm Strauss} {and} \bibinfo{person}{Juliet Corbin}.} \bibinfo{year}{1990}\natexlab{}.
\newblock \bibinfo{booktitle}{\emph{Basics of Qualitative Research: Grounded Theory Procedures and Techniques}}.
\newblock \bibinfo{publisher}{Sage Publications}, \bibinfo{address}{Newbury Park, CA}.
\newblock


\bibitem[Su and McMillan(2024)]%
        {su2024distilled}
\bibfield{author}{\bibinfo{person}{Chia-Yi Su} {and} \bibinfo{person}{Collin McMillan}.} \bibinfo{year}{2024}\natexlab{}.
\newblock \showarticletitle{Distilled GPT for source code summarization}.
\newblock \bibinfo{journal}{\emph{Automated Software Engineering}} \bibinfo{volume}{31}, \bibinfo{number}{1} (\bibinfo{year}{2024}), \bibinfo{pages}{22}.
\newblock


\bibitem[Sufi(2024)]%
        {info15020099}
\bibfield{author}{\bibinfo{person}{Fahim Sufi}.} \bibinfo{year}{2024}\natexlab{}.
\newblock \showarticletitle{Generative Pre-Trained Transformer (GPT) in Research: A Systematic Review on Data Augmentation}.
\newblock \bibinfo{journal}{\emph{Information}} \bibinfo{volume}{15}, \bibinfo{number}{2} (\bibinfo{year}{2024}).
\newblock
\showISSN{2078-2489}
\urldef\tempurl%
\url{https://doi.org/10.3390/info15020099}
\showDOI{\tempurl}


\bibitem[Tao et~al\mbox{.}(2021)]%
        {TaoWei2021Eval}
\bibfield{author}{\bibinfo{person}{Wei Tao}, \bibinfo{person}{Yanlin Wang}, \bibinfo{person}{Ensheng Shi}, \bibinfo{person}{Lun Du}, \bibinfo{person}{Shi Han}, \bibinfo{person}{Hongyu Zhang}, \bibinfo{person}{Dongmei Zhang}, {and} \bibinfo{person}{Wenqiang Zhang}.} \bibinfo{year}{2021}\natexlab{}.
\newblock \showarticletitle{On the Evaluation of Commit Message Generation Models: An Experimental Study}. In \bibinfo{booktitle}{\emph{2021 IEEE International Conference on Software Maintenance and Evolution (ICSME)}}. \bibinfo{pages}{126--136}.
\newblock
\urldef\tempurl%
\url{https://doi.org/10.1109/ICSME52107.2021.00018}
\showDOI{\tempurl}


\bibitem[Ugare et~al\mbox{.}(2024)]%
        {ugare2024improving}
\bibfield{author}{\bibinfo{person}{Shubham Ugare}, \bibinfo{person}{Tarun Suresh}, \bibinfo{person}{Hangoo Kang}, \bibinfo{person}{Sasa Misailovic}, {and} \bibinfo{person}{Gagandeep Singh}.} \bibinfo{year}{2024}\natexlab{}.
\newblock \showarticletitle{Improving llm code generation with grammar augmentation}.
\newblock \bibinfo{journal}{\emph{arXiv preprint arXiv:2403.01632}} (\bibinfo{year}{2024}).
\newblock


\bibitem[Wang et~al\mbox{.}(2024a)]%
        {wang2024trustworthyllmscodedatacentric}
\bibfield{author}{\bibinfo{person}{Chong Wang}, \bibinfo{person}{Zhenpeng Chen}, \bibinfo{person}{Tianlin Li}, \bibinfo{person}{Yilun Zhao}, {and} \bibinfo{person}{Yang Liu}.} \bibinfo{year}{2024}\natexlab{a}.
\newblock \bibinfo{title}{Towards Trustworthy LLMs for Code: A Data-Centric Synergistic Auditing Framework}.
\newblock
\newblock
\showeprint[arxiv]{2410.09048}~[cs.SE]
\urldef\tempurl%
\url{https://arxiv.org/abs/2410.09048}
\showURL{%
\tempurl}


\bibitem[Wang et~al\mbox{.}(2024c)]%
        {wang2024teachingcodellmsuse}
\bibfield{author}{\bibinfo{person}{Chong Wang}, \bibinfo{person}{Jian Zhang}, \bibinfo{person}{Yebo Feng}, \bibinfo{person}{Tianlin Li}, \bibinfo{person}{Weisong Sun}, \bibinfo{person}{Yang Liu}, {and} \bibinfo{person}{Xin Peng}.} \bibinfo{year}{2024}\natexlab{c}.
\newblock \bibinfo{title}{Teaching Code LLMs to Use Autocompletion Tools in Repository-Level Code Generation}.
\newblock
\newblock
\showeprint[arxiv]{2401.06391}~[cs.SE]
\urldef\tempurl%
\url{https://arxiv.org/abs/2401.06391}
\showURL{%
\tempurl}


\bibitem[Wang et~al\mbox{.}(2024b)]%
        {wang2024repotransbenchrealworldbenchmarkrepositorylevel}
\bibfield{author}{\bibinfo{person}{Yanli Wang}, \bibinfo{person}{Yanlin Wang}, \bibinfo{person}{Suiquan Wang}, \bibinfo{person}{Daya Guo}, \bibinfo{person}{Jiachi Chen}, \bibinfo{person}{John Grundy}, \bibinfo{person}{Xilin Liu}, \bibinfo{person}{Yuchi Ma}, \bibinfo{person}{Mingzhi Mao}, \bibinfo{person}{Hongyu Zhang}, {and} \bibinfo{person}{Zibin Zheng}.} \bibinfo{year}{2024}\natexlab{b}.
\newblock \bibinfo{title}{RepoTransBench: A Real-World Benchmark for Repository-Level Code Translation}.
\newblock
\newblock
\showeprint[arxiv]{2412.17744}~[cs.SE]
\urldef\tempurl%
\url{https://arxiv.org/abs/2412.17744}
\showURL{%
\tempurl}


\bibitem[Wolf et~al\mbox{.}(2020)]%
        {Wolf_Transformers_State-of-the-Art_Natural_2020}
\bibfield{author}{\bibinfo{person}{Thomas Wolf}, \bibinfo{person}{Lysandre Debut}, \bibinfo{person}{Victor Sanh}, \bibinfo{person}{Julien Chaumond}, \bibinfo{person}{Clement Delangue}, \bibinfo{person}{Anthony Moi}, \bibinfo{person}{Perric Cistac}, \bibinfo{person}{Clara Ma}, \bibinfo{person}{Yacine Jernite}, \bibinfo{person}{Julien Plu}, \bibinfo{person}{Canwen Xu}, \bibinfo{person}{Teven Le~Scao}, \bibinfo{person}{Sylvain Gugger}, \bibinfo{person}{Mariama Drame}, \bibinfo{person}{Quentin Lhoest}, {and} \bibinfo{person}{Alexander~M. Rush}.} \bibinfo{year}{2020}\natexlab{}.
\newblock \showarticletitle{{Transformers: State-of-the-Art Natural Language Processing}}. \bibinfo{publisher}{Association for Computational Linguistics}, \bibinfo{pages}{38--45}.
\newblock
\urldef\tempurl%
\url{https://www.aclweb.org/anthology/2020.emnlp-demos.6}
\showURL{%
\tempurl}


\bibitem[Yan et~al\mbox{.}(2023)]%
        {yan2023codetransocean}
\bibfield{author}{\bibinfo{person}{Weixiang Yan}, \bibinfo{person}{Yuchen Tian}, \bibinfo{person}{Yunzhe Li}, \bibinfo{person}{Qian Chen}, {and} \bibinfo{person}{Wen Wang}.} \bibinfo{year}{2023}\natexlab{}.
\newblock \bibinfo{title}{CodeTransOcean: A Comprehensive Multilingual Benchmark for Code Translation}.
\newblock
\newblock
\showeprint[arxiv]{2310.04951}~[cs.AI]
\urldef\tempurl%
\url{https://arxiv.org/abs/2310.04951}
\showURL{%
\tempurl}


\bibitem[Yang et~al\mbox{.}(2025b)]%
        {yang2025qwen3technicalreport}
\bibfield{author}{\bibinfo{person}{An Yang}, \bibinfo{person}{Anfeng Li}, \bibinfo{person}{Baosong Yang}, \bibinfo{person}{Beichen Zhang}, \bibinfo{person}{Binyuan Hui}, \bibinfo{person}{Bo Zheng}, \bibinfo{person}{Bowen Yu}, \bibinfo{person}{Chang Gao}, \bibinfo{person}{Chengen Huang}, \bibinfo{person}{Chenxu Lv}, \bibinfo{person}{Chujie Zheng}, \bibinfo{person}{Dayiheng Liu}, \bibinfo{person}{Fan Zhou}, \bibinfo{person}{Fei Huang}, \bibinfo{person}{Feng Hu}, \bibinfo{person}{Hao Ge}, \bibinfo{person}{Haoran Wei}, \bibinfo{person}{Huan Lin}, \bibinfo{person}{Jialong Tang}, \bibinfo{person}{Jian Yang}, \bibinfo{person}{Jianhong Tu}, \bibinfo{person}{Jianwei Zhang}, \bibinfo{person}{Jianxin Yang}, \bibinfo{person}{Jiaxi Yang}, \bibinfo{person}{Jing Zhou}, \bibinfo{person}{Jingren Zhou}, \bibinfo{person}{Junyang Lin}, \bibinfo{person}{Kai Dang}, \bibinfo{person}{Keqin Bao}, \bibinfo{person}{Kexin Yang}, \bibinfo{person}{Le Yu}, \bibinfo{person}{Lianghao Deng}, \bibinfo{person}{Mei Li}, \bibinfo{person}{Mingfeng
  Xue}, \bibinfo{person}{Mingze Li}, \bibinfo{person}{Pei Zhang}, \bibinfo{person}{Peng Wang}, \bibinfo{person}{Qin Zhu}, \bibinfo{person}{Rui Men}, \bibinfo{person}{Ruize Gao}, \bibinfo{person}{Shixuan Liu}, \bibinfo{person}{Shuang Luo}, \bibinfo{person}{Tianhao Li}, \bibinfo{person}{Tianyi Tang}, \bibinfo{person}{Wenbiao Yin}, \bibinfo{person}{Xingzhang Ren}, \bibinfo{person}{Xinyu Wang}, \bibinfo{person}{Xinyu Zhang}, \bibinfo{person}{Xuancheng Ren}, \bibinfo{person}{Yang Fan}, \bibinfo{person}{Yang Su}, \bibinfo{person}{Yichang Zhang}, \bibinfo{person}{Yinger Zhang}, \bibinfo{person}{Yu Wan}, \bibinfo{person}{Yuqiong Liu}, \bibinfo{person}{Zekun Wang}, \bibinfo{person}{Zeyu Cui}, \bibinfo{person}{Zhenru Zhang}, \bibinfo{person}{Zhipeng Zhou}, {and} \bibinfo{person}{Zihan Qiu}.} \bibinfo{year}{2025}\natexlab{b}.
\newblock \bibinfo{title}{Qwen3 Technical Report}.
\newblock
\newblock
\showeprint[arxiv]{2505.09388}~[cs.CL]
\urldef\tempurl%
\url{https://arxiv.org/abs/2505.09388}
\showURL{%
\tempurl}


\bibitem[Yang et~al\mbox{.}(2024b)]%
        {qwen2.5}
\bibfield{author}{\bibinfo{person}{An Yang}, \bibinfo{person}{Baosong Yang}, \bibinfo{person}{Beichen Zhang}, \bibinfo{person}{Binyuan Hui}, \bibinfo{person}{Bo Zheng}, \bibinfo{person}{Bowen Yu}, \bibinfo{person}{Chengyuan Li}, \bibinfo{person}{Dayiheng Liu}, \bibinfo{person}{Fei Huang}, \bibinfo{person}{Haoran Wei}, \bibinfo{person}{Huan Lin}, \bibinfo{person}{Jian Yang}, \bibinfo{person}{Jianhong Tu}, \bibinfo{person}{Jianwei Zhang}, \bibinfo{person}{Jianxin Yang}, \bibinfo{person}{Jiaxi Yang}, \bibinfo{person}{Jingren Zhou}, \bibinfo{person}{Junyang Lin}, \bibinfo{person}{Kai Dang}, \bibinfo{person}{Keming Lu}, \bibinfo{person}{Keqin Bao}, \bibinfo{person}{Kexin Yang}, \bibinfo{person}{Le Yu}, \bibinfo{person}{Mei Li}, \bibinfo{person}{Mingfeng Xue}, \bibinfo{person}{Pei Zhang}, \bibinfo{person}{Qin Zhu}, \bibinfo{person}{Rui Men}, \bibinfo{person}{Runji Lin}, \bibinfo{person}{Tianhao Li}, \bibinfo{person}{Tingyu Xia}, \bibinfo{person}{Xingzhang Ren}, \bibinfo{person}{Xuancheng Ren}, \bibinfo{person}{Yang
  Fan}, \bibinfo{person}{Yang Su}, \bibinfo{person}{Yichang Zhang}, \bibinfo{person}{Yu Wan}, \bibinfo{person}{Yuqiong Liu}, \bibinfo{person}{Zeyu Cui}, \bibinfo{person}{Zhenru Zhang}, {and} \bibinfo{person}{Zihan Qiu}.} \bibinfo{year}{2024}\natexlab{b}.
\newblock \showarticletitle{Qwen2.5 Technical Report}.
\newblock \bibinfo{journal}{\emph{arXiv preprint arXiv:2412.15115}} (\bibinfo{year}{2024}).
\newblock


\bibitem[Yang et~al\mbox{.}(2025a)]%
        {yang2025kvlinkacceleratinglargelanguage}
\bibfield{author}{\bibinfo{person}{Jingbo Yang}, \bibinfo{person}{Bairu Hou}, \bibinfo{person}{Wei Wei}, \bibinfo{person}{Yujia Bao}, {and} \bibinfo{person}{Shiyu Chang}.} \bibinfo{year}{2025}\natexlab{a}.
\newblock \bibinfo{title}{KVLink: Accelerating Large Language Models via Efficient KV Cache Reuse}.
\newblock
\newblock
\showeprint[arxiv]{2502.16002}~[cs.CL]
\urldef\tempurl%
\url{https://arxiv.org/abs/2502.16002}
\showURL{%
\tempurl}


\bibitem[Yang et~al\mbox{.}(2024a)]%
        {Yang2024Exploring}
\bibfield{author}{\bibinfo{person}{Zhen Yang}, \bibinfo{person}{Fang Liu}, \bibinfo{person}{Zhongxing Yu}, \bibinfo{person}{Jacky~Wai Keung}, \bibinfo{person}{Jia Li}, \bibinfo{person}{Shuo Liu}, \bibinfo{person}{Yifan Hong}, \bibinfo{person}{Xiaoxue Ma}, \bibinfo{person}{Zhi Jin}, {and} \bibinfo{person}{Ge Li}.} \bibinfo{year}{2024}\natexlab{a}.
\newblock \showarticletitle{Exploring and Unleashing the Power of Large Language Models in Automated Code Translation}.
\newblock \bibinfo{journal}{\emph{Proc. ACM Softw. Eng.}} \bibinfo{volume}{1}, \bibinfo{number}{FSE}, Article \bibinfo{articleno}{71} (\bibinfo{date}{July} \bibinfo{year}{2024}), \bibinfo{numpages}{24}~pages.
\newblock
\urldef\tempurl%
\url{https://doi.org/10.1145/3660778}
\showDOI{\tempurl}


\bibitem[Yu et~al\mbox{.}(2024)]%
        {yu2024codereval}
\bibfield{author}{\bibinfo{person}{Hao Yu}, \bibinfo{person}{Bo Shen}, \bibinfo{person}{Dezhi Ran}, \bibinfo{person}{Jiaxin Zhang}, \bibinfo{person}{Qi Zhang}, \bibinfo{person}{Yuchi Ma}, \bibinfo{person}{Guangtai Liang}, \bibinfo{person}{Ying Li}, \bibinfo{person}{Qianxiang Wang}, {and} \bibinfo{person}{Tao Xie}.} \bibinfo{year}{2024}\natexlab{}.
\newblock \showarticletitle{Codereval: A benchmark of pragmatic code generation with generative pre-trained models}. In \bibinfo{booktitle}{\emph{Proceedings of the 46th IEEE/ACM International Conference on Software Engineering}}. \bibinfo{pages}{1--12}.
\newblock


\bibitem[Yu et~al\mbox{.}(2023)]%
        {yu2023codereval}
\bibfield{author}{\bibinfo{person}{Hao Yu}, \bibinfo{person}{Bo Shen}, \bibinfo{person}{Dezhi Ran}, \bibinfo{person}{Jiaxin Zhang}, \bibinfo{person}{Qi Zhang}, \bibinfo{person}{Yuchi Ma}, \bibinfo{person}{Guangtai Liang}, \bibinfo{person}{Ying Li}, \bibinfo{person}{Tao Xie}, {and} \bibinfo{person}{Qianxiang Wang}.} \bibinfo{year}{2023}\natexlab{}.
\newblock \showarticletitle{CoderEval: A Benchmark of Pragmatic Code Generation with Generative Pre-trained Models}.
\newblock \bibinfo{journal}{\emph{arXiv preprint arXiv:2302.00288}} (\bibinfo{year}{2023}).
\newblock


\bibitem[Yuan et~al\mbox{.}(2023)]%
        {yuan2023evaluating}
\bibfield{author}{\bibinfo{person}{Zhiqiang Yuan}, \bibinfo{person}{Junwei Liu}, \bibinfo{person}{Qiancheng Zi}, \bibinfo{person}{Mingwei Liu}, \bibinfo{person}{Xin Peng}, {and} \bibinfo{person}{Yiling Lou}.} \bibinfo{year}{2023}\natexlab{}.
\newblock \showarticletitle{Evaluating instruction-tuned large language models on code comprehension and generation}.
\newblock \bibinfo{journal}{\emph{arXiv preprint arXiv:2308.01240}} (\bibinfo{year}{2023}).
\newblock


\bibitem[Yuan et~al\mbox{.}(2024)]%
        {YuanEvaluatingandImproving2024}
\bibfield{author}{\bibinfo{person}{Zhiqiang Yuan}, \bibinfo{person}{Mingwei Liu}, \bibinfo{person}{Shiji Ding}, \bibinfo{person}{Kaixin Wang}, \bibinfo{person}{Yixuan Chen}, \bibinfo{person}{Xin Peng}, {and} \bibinfo{person}{Yiling Lou}.} \bibinfo{year}{2024}\natexlab{}.
\newblock \showarticletitle{Evaluating and Improving ChatGPT for Unit Test Generation}.
\newblock \bibinfo{journal}{\emph{Proc. ACM Softw. Eng.}} \bibinfo{volume}{1}, \bibinfo{number}{FSE}, Article \bibinfo{articleno}{76} (\bibinfo{date}{July} \bibinfo{year}{2024}), \bibinfo{numpages}{24}~pages.
\newblock
\urldef\tempurl%
\url{https://doi.org/10.1145/3660783}
\showDOI{\tempurl}


\bibitem[Zan et~al\mbox{.}(2023)]%
        {zan2023large}
\bibfield{author}{\bibinfo{person}{Daoguang Zan}, \bibinfo{person}{Bei Chen}, \bibinfo{person}{Fengji Zhang}, \bibinfo{person}{Dianjie Lu}, \bibinfo{person}{Bingchao Wu}, \bibinfo{person}{Bei Guan}, \bibinfo{person}{Wang Yongji}, {and} \bibinfo{person}{Jian-Guang Lou}.} \bibinfo{year}{2023}\natexlab{}.
\newblock \showarticletitle{Large language models meet NL2Code: A survey}. In \bibinfo{booktitle}{\emph{Proceedings of the 61st Annual Meeting of the Association for Computational Linguistics (Volume 1: Long Papers)}}. \bibinfo{pages}{7443--7464}.
\newblock


\bibitem[Zhang et~al\mbox{.}(2023)]%
        {zhang2023draft}
\bibfield{author}{\bibinfo{person}{Jun Zhang}, \bibinfo{person}{Jue Wang}, \bibinfo{person}{Huan Li}, \bibinfo{person}{Lidan Shou}, \bibinfo{person}{Ke Chen}, \bibinfo{person}{Gang Chen}, {and} \bibinfo{person}{Sharad Mehrotra}.} \bibinfo{year}{2023}\natexlab{}.
\newblock \showarticletitle{Draft \& Verify: Lossless Large Language Model Acceleration via Self-Speculative Decoding}.
\newblock \bibinfo{journal}{\emph{arXiv preprint arXiv:2309.08168}} (\bibinfo{year}{2023}).
\newblock


\bibitem[Zhang et~al\mbox{.}(2024)]%
        {zhang2024seekerenhancingexceptionhandling}
\bibfield{author}{\bibinfo{person}{Xuanming Zhang}, \bibinfo{person}{Yuxuan Chen}, \bibinfo{person}{Yuan Yuan}, {and} \bibinfo{person}{Minlie Huang}.} \bibinfo{year}{2024}\natexlab{}.
\newblock \bibinfo{title}{Seeker: Enhancing Exception Handling in Code with LLM-based Multi-Agent Approach}.
\newblock
\newblock
\showeprint[arxiv]{2410.06949}~[cs.SE]
\urldef\tempurl%
\url{https://arxiv.org/abs/2410.06949}
\showURL{%
\tempurl}


\bibitem[Zhang et~al\mbox{.}(2025)]%
        {zhang2025llmhallucinationspracticalcode}
\bibfield{author}{\bibinfo{person}{Ziyao Zhang}, \bibinfo{person}{Yanlin Wang}, \bibinfo{person}{Chong Wang}, \bibinfo{person}{Jiachi Chen}, {and} \bibinfo{person}{Zibin Zheng}.} \bibinfo{year}{2025}\natexlab{}.
\newblock \bibinfo{title}{LLM Hallucinations in Practical Code Generation: Phenomena, Mechanism, and Mitigation}.
\newblock
\newblock
\showeprint[arxiv]{2409.20550}~[cs.SE]
\urldef\tempurl%
\url{https://arxiv.org/abs/2409.20550}
\showURL{%
\tempurl}


\bibitem[Zhong and Wang(2024)]%
        {Zhong_Wang_2024}
\bibfield{author}{\bibinfo{person}{Li Zhong} {and} \bibinfo{person}{Zilong Wang}.} \bibinfo{year}{2024}\natexlab{}.
\newblock \showarticletitle{Can LLM Replace Stack Overflow? A Study on Robustness and Reliability of Large Language Model Code Generation}.
\newblock \bibinfo{journal}{\emph{Proceedings of the AAAI Conference on Artificial Intelligence}} \bibinfo{volume}{38}, \bibinfo{number}{19} (\bibinfo{date}{Mar.} \bibinfo{year}{2024}), \bibinfo{pages}{21841--21849}.
\newblock
\urldef\tempurl%
\url{https://doi.org/10.1609/aaai.v38i19.30185}
\showDOI{\tempurl}


\end{thebibliography}

\end{document}